\documentstyle[12pt]{article}

\newtheorem{proposition}{Proposition}

\newtheorem{remark}[proposition]{Remark}
\newtheorem{definition}[proposition]{Definition}
\newtheorem{example}[proposition]{Example}

\def\+{{+\!\!\!+}}

\def\d{\partial}

\def\g{\gamma}
\def\G{\Gamma}

\def\D{{\cal D}}

\def\pmb#1{\setbox0=\hbox{#1}%
\kern.0em\copy0\kern-\wd0
\kern-.04em\copy0\kern-\wd0
\kern.08em\copy0\kern-\wd0
\kern-.04em\raise.0433em\box0 }         

\newcommand{\nc}{\newcommand}
\nc{\beq}{\begin{equation}}
\nc{\eeq}[1]{\label{#1}\end{equation}}
\nc{\ber}{\begin{eqnarray}}
\nc{\eer}[1]{\label{#1}\end{eqnarray}}
\nc{\pek}[1]{\cite{#1}}
\nc{\enr}[1]{(\ref{#1})}
\nc{\kal}[1]{{\cal{#1}}}
\nc{\dott}{\;\cdot\;}

\def\G{{\cal G}}
\def\M{{\cal M}}
\def\H{{\cal H}}
\def\g{{\mathbf g}}
\def\h{{\mathbf h}}
\def\Sy{{\cal S}}

\newcommand{\be}{\begin{equation}}
\newcommand{\ee}{\end{equation}}
\newcommand{\bea}{\begin{eqnarray}}
\newcommand{\eea}{\end{eqnarray}}

\begin{document}
\begin{center}
                                \hfill   hep-th/0311213\\
                                \hfill   LPTHE-03-30\\

\vskip .3in \noindent

\vskip .1in

{\Large \bf{Poisson sigma model over group manifolds}}
\vskip .2in

{\bf Francesco Bonechi}$^a$\footnote{e-mail address: Francesco.Bonechi@fi.infn.it}
 and  {\bf Maxim Zabzine}$^{b}$\footnote{e-mail address: zabzine@lpthe.jussieu.fr} \\


\vskip .15in

\vskip .15in
$^a${\em I.N.F.N. and Dipartimento di Fisica}\\
{\em  Via G. Sansone 1, 50019 Sesto Fiorentino - Firenze, Italy} \\
\vskip .15in
$^b${\em Lab. de Physique Th\'eorique et Hautes Energies}\\
{\em Universit\'e Pierre et Marie Curie, Paris VI}\\
{\em  4 Place Jussieu, \ 75252 Paris Cedex 05, France}

\bigskip


 \vskip .1in
\end{center}
\vskip .4in

\begin{center} {\bf ABSTRACT }
\end{center}
\begin{quotation}\noindent
 We study the Poisson sigma model which can be viewed as a topological
 string theory. Mainly we concentrate our attention on
  the Poisson sigma model over  a group manifold ${\cal G}$ with
 a Poisson-Lie structure. In this case the flat connection conditions
 arise naturally.
 The boundary conditions (D-branes) are studied in this model.
 It turns out that the D-branes are labelled by the coisotropic subgroups of ${\cal G}$.
  We give a description of the moduli space of classical solutions
 over Riemann surfaces both without and with  boundaries.
 Finally we comment briefly on the duality properties of the model.
\end{quotation}
\vfill
\eject


\section{Introduction}

The Poisson sigma model introduced in \cite{Ikeda:1993fh, Schaller:1994es} is a topological two dimensional
 field theory with the tangent space ${\cal M}$ being a Poisson manifold.
  The model is closely related to other two dimensional models such as gravity models, the Wess-Zumino-Witten
 models and two dimensional Yang-Mills theory.
  Recently the Poisson sigma model has attracted
 considerable attention due to its relation to deformation quantization. Namely it has been shown in
 \cite{Cattaneo:1999fm} that  the perturbative path integral expansion of the Poisson sigma model over the disk
 leads to the Kontsevich's star product \cite{Kontsevich:1997vb}.

In the present paper we conduct a systematic investigation of the classical Poisson sigma model
with the target space being a Poisson-Lie group. We consider the model defined over
  Riemann surfaces both with and without boundaries.
 The Poisson-Lie groups are the semiclassical limit of quantum groups,
 therefore by exploring these models at the quantum level we may hope to find  new insights
  into quantum groups. This could be considered as the main motivation for the project.
 In the present paper we take the first step in this direction and explore the classical theory,
  we hope to come back to the quantum theory elsewhere.

The key observation of the paper is that the Poisson action of a Poisson-Lie group on a target
 manifold implies the existence of a flat connection in the corresponding model.
 In particular if the target manifold is a Poisson-Lie group
  then  the on-shell Poisson sigma model can be reformulated in terms of the
 flat connections of an appropriate principal bundle and the parallel section of an associated fiber bundle.
Moreover the infinitesimal on-shell gauge transformations can be interpreted as dressing transformations
and integrated to define finite gauge transformations. This allows us to define the space of solutions
modulo gauge transformations. Since the dressing transformations are transitive on symplectic leaves, the
moduli space can be characterized in terms of the space of leaves.
Another important point is that the boundary conditions  are
 labelled by the coisotropic subgroups of the Poisson-Lie group.

 Some of these issues have been already addressed in the literature.
  Previously  the Poisson sigma model over the Poisson-Lie group has been considered in \cite{Alekseev:1995py,
  Falceto:2001eh} in connection to ${\cal G}/{\cal G}$ Wess-Zumino-Witten theories.
  While our project was in progress the work \cite{Calvo:2003kv} has
 appeared where the systematic study of Poisson sigma models over Poisson-Lie groups has been attempted.
  Despite some intersections between the results of the work \cite{Calvo:2003kv} and
 the present paper, hopefully we can offer a reasonably complete picture of the classical model
 and clarify  some important issues. The spaces of classical solutions of the Poisson sigma models
  have also been discussed previously (e.g., see the recent work \cite{Bojowald:2003pz} and the references therein).
  The important recent work \cite{Cattaneo:2003dp} should be mentioned where the first systematic study of
  general boundary conditions for the Poisson sigma model has been undertaken. In the present paper we
 clarify some general issues and as well as we give an illustration of the
 possible boundary conditions which are specific for the Poisson-Lie case.

 The paper is organized as follows. In Section~\ref{s:PLdef} we review the relevant notions from the theory
 of Poisson manifolds and Poisson-Lie groups.
  In Section~\ref{s:general} we recall the definition of the Poisson sigma model and go on to discuss
 the general boundary conditions for the model in particular. We arrive at the same result as in \cite{Cattaneo:2003dp},
 however the derivation is somewhat different. In Section~\ref{s:symmetries}
 we analyze the relation between the group action on the target space and the symmetries (and
  their generalizations) of the Poisson sigma model. The main observation is that the Poisson action of
 a Poisson-Lie group implies the flat connection conditions for the Poisson sigma model.
 Then in Section~\ref{s:group} we apply these results to the specific case when the target space is
 a group manifold itself. The on-shell  model can be rewritten in terms of new variables which have a clear geometrical
 interpretation:
 the flat connection of the principal bundle and the parallel section of the associated fiber
 bundle. We also offer the appropriate description of the boundary conditions in this context.
Using these results in Section~\ref{s:moduli} we construct the moduli spaces of the classical
  solutions of the model over a generic Riemann surface both with and without boundaries.
The description that we obtain connects the moduli space to the space of symplectic leaves.
 This space of leaves describes
a very intrinsic property of the Poisson structure. Since our considerations are only based on
 the condition that the dressing transformations are complete, the results are very general. On the other hand
it is necessary to have  more specific information about the model in order to give more
  explicit description of the moduli space.
 As an illustration of our formal results we discuss briefly the BF-theory in Section~\ref{s:bf}.
 Section~\ref{s:duality} contains some observations about the duality which is supposed to relate two
 models over different (but dual) Poisson-Lie groups.
Finally, in Section~\ref{s:end} we summarize the
 results and offer some speculations about the possible further development of our work.

\section{Poisson structures associated to Lie groups}
\label{s:PLdef}

In this Section we review some basic notions and fix notations.
Namely we collect some general facts concerning Poisson
manifolds and Poisson-Lie groups, see \cite{Va} and \cite{YJS} for a general reference.

A smooth manifold $\M$ is called a {\it Poisson}
manifold if there exists a tensor $\alpha\in \bigwedge^2T^*\M$ such
that $[\alpha,\alpha]_{sn}=0$. The bracket $[\,\,,\,\,]_{sn}$ denotes the
 the Schouten-Nijenhuis bracket for the antisymmetric contravariant tensor fields.
 In local coordinates,
$\alpha=\alpha^{\mu\nu}\partial_\mu\wedge\partial_\nu$, this amounts to the following equation
\beq
\alpha^{\mu\rho}\partial_\rho\alpha^{\nu\sigma} +
\alpha^{\sigma\rho}\partial_\rho\alpha^{\mu\nu} +
\alpha^{\nu\rho}\partial_\rho\alpha^{\sigma\mu} = 0 \;.
\eeq{defpoi}
The Poisson bracket on $C^\infty(\M)$ is defined as $\{f,g\} =
\langle df\otimes dg,\alpha\rangle$. A map $\phi:
\M\rightarrow{\cal N}$ between two Poisson manifolds is a Poisson
map if $\phi_*\alpha_\M(x)=\alpha_{\cal N}(\phi(x))$, for each
$x\in \M$.

If $\M$ is an even dimensional manifold and $\alpha$ has maximal rank then
$\alpha^{-1}$ is a symplectic form. In the general case $\alpha$
defines a symplectic foliation. The tangent space to the symplectic leaf
passing through $x\in\M$ is $\sharp(T^*_x\M)$ where the sharp map
$\sharp: T^*\M_x\rightarrow T\M_x$ is defined by $\sharp(\omega_x)
= \langle \omega_x,\alpha(x)\rangle$, for $\omega_x\in T^*_x\M$. In the local coordinates
we have that $\sharp(dx^\mu)=\alpha^{\mu\nu}\partial_\nu$. Each leaf turns out to be symplectic.
A Poisson manifold can be defined by giving its symplectic
 foliation instead of the Poisson bivector.

One can define a Lie-bracket on one forms $\Omega^1(\M)$; it satisfies
$\{df,dg\}=d\{f,g\}$ for every $f,g\in C^\infty(\M)$ and $\sharp$ defines a Lie algebra
homomorphism between forms and vector fields. We refer to \cite{Va}
for the complete definition.

 A submanifold ${\cal D}$ of $\cal M$ is called {\it coisotropic} if
$\sharp(N^*_x{\cal D})\subset T_x {\cal D}$ for each $x\in{\cal
D}$, where $N^*_x{\cal D}=\{\omega_x\in T^*_x\M \,|\, \langle
\omega_x, v_x\rangle = 0, \, \forall\ v_x\in T_x{\cal D} \}$ is
the fibre in $x\in{\cal D}$ of the conormal bundle of $\cal D$.
Symplectic leaves and preimages of symplectic leaves under a
Poisson map are examples of coisotropic submanifolds.

If the Poisson manifold is a Lie group $\G$ then it is natural to
ask about the  compatibility between the Poisson bracket and the group
  multiplication.
A Lie group $\cal G$ which is also a Poisson manifold is called a
{\it Poisson-Lie group} if the group multiplication $m: \G\times
\G\rightarrow \G$ is a Poisson map (where we assume on
$\G\times \G$ the canonical Poisson structure of the direct
product). Let us denote by $\ell_g$ and $r_g$ the left and right
translations of $g \in \G$ correspondingly. Then one can show that $\G$ is a
Poisson-Lie if and only if the Poisson tensor $\alpha$ satisfies
 the following property
\beq
\alpha(g_1g_2) = \ell_{g_1*}(\alpha(g_2)) + r_{g_2*}(\alpha(g_1))
\;.
\eeq{PLdefbasic}
 One can verify that $\alpha(e)=0$, {\it i.e.} the identity
is always a degenerate point, so that a Poisson-Lie group can
never be symplectic.

A crucial property for a simply connected Poisson-Lie group is
that it defines a dual Poisson-Lie group. Let $\mathbf g$ be the
Lie algebra of $\G$; on ${\mathbf g}^*\equiv T^*_e\G$ the bracket
\beq
\{d_ef,d_eg\}=d_e\{f,g\},\,\,\,\,\,\,\,\,\,\,\,\,\,f,g\in C^\infty(\G)
\eeq{brform}
 defines a
structure of Lie algebra. Let $\G^*$ be the (simply connected)
group whose Lie algebra is ${\mathbf g}^*$. It can be shown that
also $\G^*$ is a Poisson-Lie group and that $\G^{**}$ is $\cal G$;
we refer to $\G^*$ as the {\it dual group} of $\G$.

The left invariant forms in $\G$ close a finite dimensional subalgebra
of $\Omega^1(\G)$ and define a canonical realization of
$\g^*$; by applying the $\sharp$ map we also get a realization of $\g^*$
as vector fields. Let $\{T_A\}$ be a basis of $\g$ and let $\{T^A\}$
be the dual basis of $\g^*$: structure constants are defined as
$[T_A,T_B]= f_{AB}^{\,\,\,\,\,\,\,\,\,C} T_C$ and
  $[T^A,T^B]= \tilde{f}^{AB}_{\,\,\,\,\,\,\,\,\,C} T^C$. Let
$\{\omega^A\}$ be the corresponding basis of left invariant forms on $\G$,
and let $\{s^A= \sharp\omega^A\}$ be the so called {\it dressing
vector fields}. Thus we have that
\beq
[ s^A,s^B ] = \tilde{f}^{AB}_{\,\,\,\,\,\,\,\,\,C} s^C,
\eeq{dresvectf}
 where $[\,\,,\,\,]$ is the Lie bracket for the vector fields and
\beq
\{\omega^A,\omega^B\} = \tilde{f}^{AB}_{\,\,\,\,\,\,\,\,\,C} \omega^C  \;.
\eeq{fo12}
If the dressing vector fields are complete then the corresponding
action of $\G^*$ on $\G$ is called the {\it dressing action} and $\G$ is called
a {\it complete} Poisson-Lie group. For example, any compact Poisson-Lie group and
 its dual are complete.

The action of a Poisson-Lie group $(\G, w)$ on a Poisson manifold
$(\M, \alpha)$ is Poisson if the map $\phi:
\G\times\M\rightarrow\M$, $\phi(g,x)=gx$ is a Poisson map. One can
show that this leads to the following property
\begin{equation}\label{poisson_action}
\alpha (gx) = \phi_{g*}\alpha (x) + \phi_*^x w(g) \;,
\end{equation}
where $\phi_g:\M\rightarrow\M$, $\phi_g(x) = \phi(g,x)$, and
$\phi^x:\G\rightarrow\M$, $\phi^x(g)=\phi(g,x)$. The dressing
action of $\G^*$ on $\G$ is Poisson; furthermore it is transitive
on symplectic leaves.  The symplectic leaves are the orbits of the dressing action
 and therefore, in analogy with the coadjoint orbits, the symplectic leaf
 can be thought of as a homogeneous space ${\cal S} \sim \G^*/\H_{({\cal S}, x_0)}$, where
 $\H_{({\cal S},x_0)}$ is stability subgroup of a fixed point $x_0 \in {\cal S}$.

A subgroup $\H$ is a {\it Poisson-Lie subgroup} if it is a Poisson-Lie group itself and the
injection map is a Poisson morphism.
A subgroup $\H\subset\G$ is said to be {\it coisotropic} if it is
a coisotropic submanifold (see \cite{STS,Weinstein,Lu-thesis} for what follows).
Every Poisson-Lie subgroup is coisotropic but it is not true the opposite.
Let $\H$ be connected and let $\h$ be its Lie algebra.
It can be shown that $\H\subset\G$ is coisotropic if and only if
${\mathbf h}^\perp$ is a Lie subalgebra of ${\mathbf g}^*$, where ${\mathbf h}^\perp\subset{\mathbf g}^*$
is the annihilator of ${\mathbf h}$.
To each
coisotropic subgroup $\H$ of $\G$ we can associate the subgroup
$\H^\perp$ of $\G^*$, whose Lie algebra is ${\mathbf h}^\perp$.
Since ${{\mathbf h}^\perp}^\perp={\mathbf h}$ then $\H^\perp$ is a
coisotropic subgroup of $\G^*$; we call $\H^\perp$ the {\it
complementary dual} of $\cal H$.

If $\H$ is a closed coisotropic subgroup, the map
$p_\H:\G\rightarrow\G/\H$ defines a Poisson tensor
on the homogeneous space $\G/\H$ such that $p_\H$ is a Poisson map. So the action of
$\G$ on $\G/\H$ is Poisson and
$p_\H(e)\in\G/\H$ is a degenerate point; in this case
the Poisson structure is never
symplectic.

We will need the following observation. Let us introduce a basis in $\g$
such that $\{T_A\}_{A=1}^{n_{\H}}$ is a basis for $\h$, it is clear that $\{\omega^A(x)\}_{A>n_\H}$
is a basis for $N_x^*\H$ for each $x\in\H$ and $s^A(x)\in T_x\H$
for each $A>n_\H$; {\it i.e.} the dressing action of $\H^\perp$
leaves $\H$ invariant.

The important role played by coisotropic subgroups is fully appreciated in the quantization
process (see \cite{Ciccoli} for the following considerations).
There exists a quantum counterpart of the construction of Poisson
homogeneous spaces described above. It is an experimental fact of quantum groups that
coisotropic subgroups are quantized; they
define the so called {\it quantum coisotropic subgroups} of quantum groups. By using
them it is possible to construct all the so called embeddable quantum homogeneous spaces as quotient
of quantum groups. See for instance \cite{Bonechi:2000ia} for the role played
by coisotropic subgroups of standard $U(4)$ in defining a
quantum version of the instanton bundle.

Finally let us recall the definition of the {\it Drinfeld double}
${\mathbf D}(\g)$. Let ${\mathbf D}(\g)=\g\oplus\g^*$ be the
direct sum of vector spaces. Then there exists a unique Lie
algebra structure on ${\mathbf D}(\g)$ such that $\g$ and $\g^*$
are Lie subalgebras and isotropic with respect to the degenerate
pairing $\langle X_1+Y_1,X_2+Y_2\rangle=Y_1(X_2)+Y_2(X_1)$.

\section{Poisson sigma model}
\label{s:general}

Let us start by introducing the Poisson sigma model. The model has two bosonic
 real fields, $X$ and $\eta$. $X$ is a map\footnote{In Section \ref{s:group}
 we will discuss that sometimes
 $X$ can be treated  as a section of an appropriate fibre bundle.} from the two-dimensional worldsheet
 $\Sigma$ (possibly with boundaries) to a Poisson manifold $\M$ of dimension $d$ and $\eta$ is
 a differential form on $\Sigma$ taking values in the pull-back by $X$ of the cotangent
 bundle of $\M$, i.e. a section of $X^*(T^*\M) \otimes T^* \Sigma$. In local coordinates, $X$ is given by
$d$ functions $X^\mu(\xi)$ and $\eta$ by $d$ differential
 1-forms $\eta_\mu(\xi)=\eta_{\alpha\mu} d\xi^{\alpha}$ (where $\alpha, \beta = 1,2$ and $\mu,
  \nu = 1, ..., d$). The action of the Poisson sigma model
 given by the following expression
\beq
 S= \int\limits_{\Sigma}
  d^2\xi \,\left [ \epsilon^{\alpha\beta} \eta_{\alpha \mu} \d_\beta X^\mu
 + \frac{1}{2} \alpha^{\mu\nu}(X) \eta_{\alpha\mu} \eta_{\beta\nu} \epsilon^{\alpha\beta} \right ] ,
\eeq{actionps}
 where  $\alpha^{\mu\nu}$ is a Poisson tensor on $\M$.
 The variation of the action gives rise to the following equations of
 motion
\beq
 d\eta_\rho + \frac{1}{2} (\d_\rho \alpha^{\mu\nu}) \eta_\mu \wedge \eta_\nu =0,\,\,\,\,\,\,\,\,\,\,\,
 dX^\mu + \alpha^{\mu\nu}\eta_\nu = 0 ,
\eeq{eqmotion}
 and  if $\d\Sigma \neq \emptyset $
then we should impose the boundary condition such that
\beq
   (\eta_{\tau\mu} \delta X^\mu)|_{\d\Sigma}=0 ,
\eeq{boundcond}
 where $\eta|_{T^*\d\Sigma}=\eta_{\tau\mu}d\tau$.
 The action (\ref{actionps}) is invariant under the infinitesimal gauge transformations
\beq
 \delta_\beta X^\mu = \alpha^{\mu\nu} \beta_\nu,\,\,\,\,\,\,\,\,\,\,\,\,\,
 \delta_\beta \eta_\mu = - d\beta_{\mu} - (\d_\mu \alpha^{\nu\rho}) \eta_\nu \beta_\rho ,
\eeq{gaugetransf}
  with the parameter $\beta_\mu dX^\mu \in \Gamma(X^*(T^*\M))$,
  where $\Gamma$ denotes the space of sections,
such that the following boundary condition is satisfied \beq
 (\beta_\mu \d_\tau X^\mu)|_{\d\Sigma}=0 .
\eeq{gaugebc}

 Now let us describe the boundary conditions which satisfy (\ref{boundcond}) and (\ref{gaugebc}). Locally
 it is clear that along the boundary $\d \Sigma$ in some direction $X$ should be taken of fixed value and
 in the remaining directions $\eta_\tau$ and gauge parameter $\beta$  should be taken zero. Thus in the
 covariant language the boundary condition is given by the requirement  that
\beq
   X : \d \Sigma \longrightarrow {\cal D} \subset \M,
\eeq{bcglobp}
 i.e. the image of the boundary is a submanifold ${\cal D}$ of $\M$ and
\beq
 \eta|_{ T^* \d \Sigma} \in \Gamma(X^*(N^* {\cal D})) ,\,\,\,\,\,\,\,\,\,\,\,
 \beta|_{\d\Sigma} \in \Gamma(X^*(N^* {\cal D}))
\eeq{betaetabc}
 and where $N^* {\cal D}$ is the conormal bundle of ${\cal D}$.
 Here we assume that $\d \Sigma$
 has a single component. This description of boundary conditions is very much analogous to
  the boundary conditions of the open string theory known as D-branes. Hence we may borrow the string
 jargon and refer to the submanifold ${\cal D}$ as D-brane.

 Indeed as it stands the boundary conditions (\ref{bcglobp}) and (\ref{betaetabc}) are not invariant
 under the residual gauge transformations (i.e., the gauge transformations (\ref{gaugetransf})
 restricted to the boundary $\d\Sigma$ using the boundary conditions (\ref{bcglobp}) and (\ref{betaetabc})).
 To illustrate the point let us use the local coordinates.
 In the neighborhood of a point $x_0 \in {\cal D}$ we
  choose the coordinates $X^\mu = (X^a, X^n)$ adapted to the submanifold ${\cal D}$ such that in this
neighborhood the  submanifold ${\cal D}$ is given by the condition $X^a=0$.
  We use the Latin lower case letter from the beginning  of alphabet for the
 coordinates transverse to the submanifold  ${\cal D}$  and from the middle for the coordinates
 along the submanifold ${\cal D}$. In these coordinates the condition (\ref{betaetabc}) becomes
   $\eta_{\tau n} =0$ and $\beta_n=0$.
 The gauge transformations restricted to the boundary should leave the boundary conditions invariant.
 Namely the following should be true
\beq
 \delta_\beta X^a|_{\d\Sigma}= \alpha^{a\mu}\beta_\mu|_{\d \Sigma} = \alpha^{ab} \beta_b |_{\d\Sigma} =0,
\eeq{invabc}
\beq
  \delta_\beta \eta_{\tau n}|_{\d\Sigma} =  - \d_n \alpha^{ab} \eta_{\tau a} \beta_b =0.
\eeq{taugaugebc}
 Since $\beta_b$ and $\eta_{\tau a}$ are unrestricted along the boundary we have to impose
 that $\alpha^{ab}(0,X^n)=0$ (as a result of this $\d_n \alpha^{ab}(0,X^n) = 0$).
 Therefore $\sharp N^*{\cal D}\subset T{\cal D}$ and the submanifold ${\cal D}$ is coisotropic
with respect to the Poisson structure
 $\alpha$. We have also to check that
 the remaining gauge algebra acting on unrestricted fields closes on the boundary (at least on-shell).
 Indeed using the property of coisotropy,  $\alpha^{ab}(0,X^n)=0$, we obtain the (on-shell) closure
\beq
  [\delta_{\beta_2}, \delta_{\beta_1}] X^n = \alpha^{n\mu} (\d_\mu \alpha^{ab} \beta_{2b}\beta_{1a}) =
 \alpha^{nc} (\d_c \alpha^{ab} \beta_{2b}\beta_{1a}) ,
\eeq{Xtranbound}
\beq
 [\delta_{\beta_2}, \delta_{\beta_1}] \eta_{\tau a} =\delta_{(\d_c \alpha^{ab} \beta_{2b}\beta_{1a})} \eta_{\tau a} ,
\eeq{etatranboundcl}
 where in the last equation we have used the equation of motion restricted to the boundary.

 Thus we can conclude that the boundary conditions for the Poisson
 sigma model are given by (\ref{bcglobp}) and (\ref{betaetabc})
 where the submanifold ${\cal D}$ is coisotropic with respect to $\alpha$.
 The coisotropy follows from the consideration of the gauge symmetry on the boundary. This coincides
with the results of \cite{Cattaneo:2003dp}.

 However we would like to bring the reader attention to a  subtlety in the derivation of the boundary conditions.
 In some instances it is more convenient to write the action of the model in a form which
 differs from (\ref{actionps}) by a boundary term. For example, for the case of the linear Poisson
 structure the action is typically written in the following form
\beq
 S = \int\limits_{\Sigma}
  d^2\xi \,\,X^A\left (  \epsilon^{\alpha\beta} \d_\alpha \eta_{\beta A}
 + \frac{1}{2} \tilde{f}^{BC}_{\,\,\,\,\,\,\,\,\,A} \eta_{\alpha B} \eta_{\beta C} \epsilon^{\alpha\beta} \right ) ,
\eeq{BFtheacdef}
 where $\M$ is a vector space, $\tilde{f}^{BC}_{\,\,\,\,\,\,\,\,\,A}$ are  the structure constants
 of a Lie algebra
 and the Poisson structure $\alpha^{AB} =\tilde{f}^{AB}_{\,\,\,\,\,\,\,\,\,C} X^C$.
 The action (\ref{BFtheacdef}) is the standard form for the BF-theory (for further details see Section \ref{s:bf}).
 The action (\ref{BFtheacdef}) differs from
 (\ref{actionps}) by a boundary term. Infact the action (\ref{BFtheacdef}) is invariant under gauge
transformations without any restriction on the boundary. Nevertheless one can still repeat the analysis
coming from the requirements (\ref{invabc})-(\ref{etatranboundcl}) and arrive at the same result.
Namely, the boundary conditions should be given by (\ref{bcglobp}) and (\ref{betaetabc}) and
${\cal D}$ is coisotropic submanifold of $\M$. The restriction of the gauge transformations on the boundary
come now from the requirement that they close an algebra and leave the boundary conditions invariant.

\section{Symmetries of Poisson sigma model}
\label{s:symmetries}

In this Section we consider the global symmetry associated with
group action on $\M$ and its
  generalization for the Poisson sigma model.

 Let us assume that there exist an action of a Lie group ${\cal G}$ on $\M$
\beq
 \phi:  {\cal G} \times \M \longrightarrow \M .
\eeq{defaction}
 At the infinitesimal level this action is realized by the vector fields $k_A=k^\mu_A\partial_\mu$ which
 obey the corresponding Lie algebra ${\mathbf g}$ relations
\beq
 [ k_A, k_B ] = f_{AB}^{\,\,\,\,\,\,\,\,\,C} k_C .
\eeq{alegbra}
 Under the action of the group $\G$  the fields of
the model have
 the following transformations
\beq
 \delta X^\mu = a^A k^\mu_A(X),\,\,\,\,\,\,\,\,\,\,\,\,\,\,\,
 \delta \eta_{\alpha\mu} = - a^A \eta_{\alpha\nu} k^\nu_{A,\mu} ,
\eeq{tranfields}
 where $k^\nu_{A,\mu} \equiv \d_\mu k^\nu_A$ and $a^A$ being the parameter of
 the transformations. The variation of the action (\ref{actionps})
 under (\ref{tranfields}) is given by
\beq
 \delta S = \frac{1}{2}\int d^2 \xi \, \epsilon^{\alpha \beta} a^A
 \eta_{\alpha\mu} \eta_{\beta\nu} \left ({\cal L}_{k_A} \alpha^{\mu\nu}\right ) +
 \int d^2 \xi \,\epsilon^{\alpha\beta} \d_\beta a^A (\eta_{\alpha\mu} k^\mu_A) .
\eeq{varact}
 If ${\cal L}_{k_A} \alpha^{\mu\nu} =0$ then the transformations (\ref{tranfields})
 are the global symmetries of the action. Thus in this case the Poisson structure $\alpha$ is
 invariant under the action of $\phi$,
\beq
 \alpha (gm) = \phi_{g*} \alpha(m),\,\,\,\,\,\,\,\,\forall g\in \G,\,\,\forall m\in \M .
\eeq{actinvar}
 For this symmetry the corresponding conserved
 current is $J_{\alpha A} = \eta_{\alpha\mu} k^\mu_A$ with the conservation law
\beq
 d J_A = \epsilon^{\alpha\beta} \d_\alpha J_{\beta A} = 0.
\eeq{conservabel}
 The gauge transformations (\ref{gaugetransf}) imply the following transformations for $J_A$
\beq
 \delta_\beta J_A = - d(\beta_\mu k^\mu_A) + \beta_\mu k^\mu_{A,\rho} (dX^\rho +\alpha^{\rho\sigma}\eta_\sigma) ,
\eeq{gaugetransfabel}
 where the last term is proportional to the equation of motion. Hence we conclude that for
  the Poisson sigma model with the $\G$-invariant Poisson structure  we can construct (on-shell)
 $(\dim{\mathbf g})$ abelian flat connections, which may, however,
  be dependent.

 The above situation can be generalized in the following way. Let us assume that the group $\G$
 is a Poisson-Lie group with $w$ being the corresponding Poisson-Lie tensor.
 Then it is natural to consider that the action $\phi$ is a  Poisson action.
If $\G$ is connected, at the infinitesimal level the condition (\ref{poisson_action}) becomes
\beq
 {\cal L}_{k_A} \alpha^{\mu\nu} =\tilde{f}^{CB}_{\,\,\,\,\,\,\,\,\,A} k_C^\mu k_B^\nu ,
\eeq{condonalpha}
 where $\tilde{f}^{CB}_{\,\,\,\,\,\,\,\,\,A}$ are the structure constants on the dual Lie algebra
  ${\mathbf g^*}$. Defining   the current $J_A = k_A^\mu \eta_\mu$ as before we can show that the
 equation of motion (\ref{eqmotion}) together with the property (\ref{condonalpha}) imply
\beq
  d J_A +  \frac{1}{2} \tilde{f}^{CB}_{\,\,\,\,\,\,\,\,\,A}\, J_C \wedge J_B = 0 .
\eeq{connonable}
 In its turn the gauge transformations (\ref{gaugetransf}) lead to the following transformations of $J_A$
\beq
 \delta_\beta J_A = - d(\beta_\mu k^\mu_A) - \tilde{f}^{CB}_{\,\,\,\,\,\,\,\,\,A} J_C (\beta_\rho k^\rho_B)
 + \beta_\mu k^\mu_{A,\sigma} (dX^\sigma +\alpha^{\sigma\nu}\eta_\nu) ,
\eeq{gauegrpl}
 where again the last term is proportional to the equations of motion. Thus on-shell  $J=J_A T^A$
 can be interpreted as a flat connection of the trivial principal bundle $\Sigma\times\G^*$, where $\G^*$ is
 the simply connected group corresponding to the dual Lie algebra ${\mathbf g}^*$.
 Following the Klim\v{c}ik-Severa \cite{Klimcik:1995ux} we can understand the flatness condition
 (\ref{connonable}) as a sort of generalization of  the conservation
 law; we call it the Poisson-Lie symmetry.
 Locally the equation (\ref{connonable}) can be solved easily $J=\tilde{g}^{-1} d\tilde{g}$
 where $\tilde{g}$ is a map  from the worldsheet $\Sigma$ to the group ${\cal G^*}$.

Indeed the case of invariant Poisson structure can be embedded to the Poisson-Lie framework assuming
 the trivial Poisson-Lie structure, i.e. $w=0$.

\section{Poisson sigma model over group manifold}
\label{s:group}

In this Section we consider the case when the target space $\M$ can be identified with a Poisson-Lie group $\G$.
The multiplicative property (\ref{PLdefbasic}) of the Poisson tensor $\alpha$ implies
that, for instance, the left multiplication is a Poisson action of $\G$ on itself, see (\ref{poisson_action})
with $w=\alpha$. We can apply the considerations of Poisson-Lie symmetry of the previous section and define the
current $J_A$ associated to the left multiplication, {\it i.e.} $J_A=k_A^\mu\eta_\mu$, where $k_A=k_A^\mu\partial_\mu$
are now the left invariant vector fields on $\G$.

The important issue now is that $\omega^A_\mu$ is inverse of $k^\mu_A$ and the following two sets
of variables are equivalent
\beq
 (X^\mu, \eta_\mu, \beta_\mu)\,\,\,\,\,\Longleftrightarrow \,\,\,\,\,(X^\mu, J_A =k_A^\mu \eta_\mu,
 \beta_A = - k_A^\mu\beta_\mu).
\eeq{newvar}
The equations of motion in the new variables $(X,J)$ read
\beq
 dX^\mu - s^{A\mu} J_A =0,\,\,\,\,\,\,\,\,\,\,\,\,\,\,\,
 d J_A +  \frac{1}{2} \tilde{f}^{CB}_{\,\,\,\,\,\,\,\,\,A}\, J_C \wedge J_B = 0
\eeq{neweqofmot}
and the on-shell gauge transformations are
\beq
 \delta_\beta X^\mu = s^{A\mu} \beta_A,
\eeq{newgautran1}
\beq
  \delta_\beta J_A = d \beta_A + \tilde{f}^{CB}_{\,\,\,\,\,\,\,\,\,A} J_C \beta_B
\eeq{newgautran}
 where $s^{A\mu}= - \alpha^{\mu\nu} \omega^A_\nu$. One can easily check that
 the equations (\ref{neweqofmot}) and the transformations (\ref{newgautran1}) and
 (\ref{newgautran})
 are {\em on-shell completely equivalent} to the equations (\ref{eqmotion}) and the transformations (\ref{gaugetransf})
 taking into account the definition (\ref{newvar}).

The vector fields $s^A=s^{A\mu}\partial_\mu=\omega^A_\nu\alpha^{\nu\mu}\partial_\mu=\sharp\omega^A$
are the dressing vector fields described in Section \ref{s:PLdef}; they satisfy (\ref{dresvectf}).
The dressing action is a Poisson action and thus we have
\beq
 {\cal L}_{s^A} \alpha^{\mu\nu} = f_{CB}^{\,\,\,\,\,\,\,\,\,A} s^{C\mu} s^{B\nu},
\eeq{dresgrSS}
which can be explicitly checked. We can now apply the discussion of the
 Poisson-Lie symmetry from Section \ref{s:symmetries},
this time with respect to the Poisson
action given by the dressing action of $\G^*$ defined by $s^A$.
We then introduce the current $j^A = s^{A\mu}\eta_\mu$; as a consequence of the Poisson symmetry of the dressing action,
it satisfies the flat condition
\beq
 d j^A + \frac{1}{2} f_{BC}^{\,\,\,\,\,\,\,\,\,A} j^B \wedge j^C = 0
\eeq{CartMa}
 and transforms under the on-shell gauge symmetries as follows
\beq
\delta_\beta j^A =  d\tilde{\beta}^A + f_{CB}^{\,\,\,\,\,\,\,\,\,A} j^C \tilde{\beta}^B ,
\eeq{gautrandual}
 where $\tilde{\beta}^A=  -\beta_\mu s^{A\mu}$.
 However it is interesting to note that on-shell $j^A = - \omega^A_\mu dX^\mu$.
 Now if we adopt this as the definition of $j^A$ the condition (\ref{CartMa}) is satisfied  trivially
 and the gauge transformations (\ref{gautrandual}) of $j^A$ are off-shell.

 The infinitesimal on-shell gauge transformations (\ref{newgautran1}) and (\ref{newgautran}) can be easily integrated.
In fact we know that the invariant forms $\Omega^1(\G)^{\rm inv}$ close a finite dimensional subalgebra of $\Omega^1(\G)$.
If we limit the gauge transformations to $\beta:\Sigma\rightarrow\Omega^1(\G)^{\rm inv}$ then the algebra of
infinitesimal transformations is just $g(\Sigma,\g^*)=\{\beta:\Sigma\rightarrow\g^*\}$ and can be integrated to
the usual gauge group $G(\Sigma,\G^*)=\{\gamma:\Sigma\rightarrow\G^*\}$. If the Poisson-Lie group
$\G$ is complete, {\it i.e} the dressing vector fields are complete, then the gauge group acts on
the solutions $(X,J)$ of equation of motion as
a dressing transformation on $X$ and as usual transformation of connections on $J$.

We are going now to discuss the boundary conditions in the new variables. Now we specialize
to the group case the discussion of the boundary conditions given in Section \ref{s:general} for the model on a generic
Poisson manifold. There we saw that we have to require that $X$ maps
the boundary $\d\Sigma$ in a coisotropic submanifold $\D\subset\G$, {\it i.e.}
$X : \d \Sigma \longrightarrow {\cal D}$.
The boundary term (\ref{boundcond}) can be written as follows
\beq
  (\eta_{\tau\mu}\delta X^\mu)|_{\d\Sigma}=(\eta_{\tau\mu} k^\mu_A \omega^A_\nu \delta X^\nu)|_{\d\Sigma}=
 - (J_{A\tau} j^A_\tau)|_{\d\Sigma}=0 \;.
\eeq{boundcondgman}
In terms of the currents $J$ and $j$, the boundary conditions
can be imposed as follows
\beq
 J|_{ T^* \d \Sigma} \in \Omega^1(\Sigma)\otimes{\mathbf h}^{\perp} \subset \Omega^1(\Sigma)\otimes{\mathbf g}^* ,\,\,\,\,\,\,\,\,\,\,\,
 j|_{ T^* \d \Sigma} \in \Omega^1(\Sigma)\otimes{\mathbf h} \subset \Omega^1(\Sigma)\otimes{\mathbf g} ,
\eeq{jJboundarcond}
 such that
\beq
 \langle {\mathbf h}, {\mathbf h}^\perp \rangle = 0 ,
\eeq{pairbc}
where $\langle\, , \, \rangle$ is the natural pairing between ${\g}$ and ${\g}^*$
($\dim {\mathbf h} + \dim {\mathbf h}^\perp = \dim {\mathbf g}$). In order to let the
gauge transformations (\ref{newgautran}) restricted to the boundary $\d \Sigma$ close an
algebra we must ask that $\h^\perp\subset\g^*$ and $\h\subset\g$ are both subalgebras.
Let $\H^\perp\subset\G^*$ and $\H\subset \G$ be the connected subgroups whose Lie algebras are respectively
$\h^\perp$ and $\h$; we know from
Section \ref{s:PLdef} that this condition is equivalent to saying that $\H\subset\G$ and $\H^\perp\subset\G^*$
are coisotropic subgroups and are complementary duals of each other. Finally, the underlying submanifold $\D$
is both $\H$ and $\H^\perp$ invariant and coisotropic.

To clarify the discussion we can analyze the boundary condition in local coordinates. As it has been
 discussed in Section \ref{s:general} in local coordinates the boundary conditions are given by
\beq
 \eta_{\tau n}|_{\d\Sigma} =0,\,\,\,n=1, ...,\dim {\cal D},\,\,\,\,\,\,\,\,\,\,\,\,\,
 X^a|_{\d\Sigma}=0,\,\,\,a= \dim{\cal D}+1, ..., \dim {G}.
\eeq{locbcXeta}
According to (\ref{boundcondgman})-(\ref{pairbc}) we want to impose the following
 boundary conditions on the currents
\beq
 J_{\tau A}|_{\d\Sigma}=0,\,\,\,A = 1, ...,\dim {\mathbf h},\,\,\,\,\,\,\,\,\,\,\,\,\,
 j^A_\tau |_{\d\Sigma}=0,\,\,\,A= \dim{\mathbf h}+1, ..., \dim {\mathbf g} .
\eeq{boundcjjjj}
 Together the conditions (\ref{locbcXeta}) and (\ref{boundcjjjj})
imply the following conditions on the invariant vector fields  $k_A$ and
 the invariant one forms $\omega^A$
\beq
 k^a_{A}|_{\cal D} =0\,\,\,A = 1, ...,\dim {\mathbf h},\,\,\,\,\,\,\,\,\,\,\,\,\,
 \omega^{A}_n|_{\cal D} =0\,\,\,A = \dim{\mathbf h} +1, ..., \dim {\mathbf g},
\eeq{kkkkvetc}
 where we have used the fact that $X^n$ and $\eta_{\tau a}$ are unrestricted on the boundary.
 In the covariant language the condition (\ref{kkkkvetc}) becomes
\beq
\{ k_{A} \}_{A=1}^{\dim {\mathbf h}}|_{\cal D} \subset T{\cal D},
\,\,\,\,\,\,\,\,\,\,\,\,\,\,\,\,\,
 \{ \omega^{A} \}_{A={\mathbf h} +1}^{\dim {\mathbf g}}|_{\cal D} \subset N^*{\cal D}.
\eeq{condcovKKK}
Since we are dealing with a group manifold the vector fields $\{k_A\}$ are linearly independent at each point.
Thus due to this fact and  the property (\ref{condcovKKK}) the manifold ${\cal D}$ has
dimension equal to $\dim {\mathbf h}$. The set of vector fields  $\{ k_{A} \}_{A=1}^{\dim {\mathbf h}}|_{\cal D}$
form a basis in the tangent space of the manifold ${\cal D}$ and therefore must be in
involution. We can conclude that ${\mathbf h}$ is a subalgebra.
In its turn the coisotropy of ${\cal D}$ implies that
\beq
\{ k_{A} \}_{A=1}^{\dim {\mathbf h}}|_{\cal D} \subset T{\cal D},
\,\,\,\,\,\,\,\,\,\,\,\,\,\,\,\,\,
 \{ s^{A} \}_{A={\mathbf h} +1}^{\dim {\mathbf g}}|_{\cal D} \subset T{\cal D},
\eeq{condcovKKKSSS}
with $s^A$ being the dressing vector fields.
As it was stated in the Section \ref{s:general} the coisotropy of ${\cal D}$ is essential to guarantee that
the gauge symmetry restricted to the boundary is consistent, i.e. it leaves the boundary conditions
invariant and it closes to an algebra (at least on-shell) on the boundary.

Now the new point is that in the case of a Poisson-Lie group, the model
can be written in new variables and it makes sense to consider the on-shell gauge transformations
as maps from $\Sigma$ to the dual group $\G^*$. Therefore in this context it is natural to require that the symmetry
on the boundary have the same nature as in the bulk. Namely we have to require that
${\mathbf h}^\perp$ forms a subalgebra of ${\mathbf g}^*$
and thus the gauge transformations (\ref{newgautran}) have a natural
restriction to the boundary. In this case $J$ and $j$ (on-shell)
 transform as gauge connections both in the bulk and on the boundary
and their algebra of gauge transformations has field independent structure constants.
Once that we ask that  ${\mathbf h}$ and ${\mathbf h}^\perp$ are Lie subalgebras, then the submanifold ${\cal D}$
is invariant under the action of $\H$ and $\H^\perp$ and the vectors (\ref{condcovKKKSSS}) are the fundamental
vector fields of these actions.
It is important to stress that not any subgroup $\H$ of $\G$ has its dual $\H^\perp
\subset \G^*$ in the above sense. Only the coisotropic subgroups (see the definition in Section~\ref{s:PLdef})
$\H$ would satisfy this  requirement.

We finally motivated the following definition.

\medskip
\begin{definition}\label{admissible_branes_definition}
Let $\H$ be a coisotropic subgroup of $\G$. An admissible $(\H,\H^\perp)$-brane is a coisotropic submanifold $\D\subset\G$
such that it is invariant after the left action of $\H\subset\G$ and the dressing transformations of $\H^\perp$. Moreover
$\dim\D=\dim\H$.
\end{definition}

A natural candidate to be an $(\H,\H^\perp)$-brane is the subgroup
$\H$ itself. However there exists a wider class of such branes that satisfy the previous
requirements. For example, one can see the discussion of admissible branes in the context of BF-theory.

\medskip
\begin{example}{\rm
 If $\H=\G$ then $\H^\perp=\{e\}$ and the only admissible brane is $\D=\G$. In this case the boundary conditions
simply state that $J|_{\d\Sigma}=0$, without any restriction on $X$, {\it i.e.} we recover the Cattaneo and Felder
boundary conditions from \cite{Cattaneo:1999fm}.

The other extreme case is when $\H=\{e\}$ and $\H^\perp=\G^*$;
then an admissible brane is a fixed point $g_0$ of the dressing transformations of $\G^*$, {\it i.e.} a degenerate point.
In this case the boundary conditions amount to fixing $X=g_0$ on the boundary without any restriction on $J$.}
\end{example}

\section{Moduli space of solutions: group case}
\label{s:moduli}

In this Section we want to describe the moduli space of gauge
inequivalent solutions of the equations of motion for the Poisson
sigma model on a Poisson-Lie group $\G$. We assume that the
dressing vector fields are complete, so that $\G^*$ acts on $\G$
by the left dressing transformations. In this case the gauge
transformations can be integrated to finite transformations and
therefore the moduli space is well-defined. We consider the model
over an arbitrary Riemann surface $\Sigma$ either with a boundary
or without a boundary. In the case of non empty boundary the
boundary conditions should be incorporated into the task as
discussed in Section \ref{s:group}. In particular our goal is to
relate the moduli space to the structure of symplectic leaves of
$\G$. Namely we will describe it as certain union of the moduli
spaces of flat connections of those subgroups of $\G^*$ which are
stability subgroups of the symplectic leaves.

Before going into the details of the present model we need to
briefly recall some basic facts on flat connections.

\subsection{Moduli spaces of flat connections}
\label{modfl}

Starting from the Atiyah-Bott work \cite{Atiyah:fa} the moduli
space of flat connections over the Riemann surfaces has been
extensively studied in the mathematical and physical literature.
The typical description of the moduli space of flat $\H$
connections over $\Sigma$, with $\H$ being a Lie group, is
$Hom(\pi_1(\Sigma), \H)/Ad\, \H$, which in the case of
two-dimensional surfaces is a symplectic space. We are mainly
interested to focus on  the flat structures that lie behind this
description. This approach has been advocated in
\cite{Ivanova:1999iq} and, since it will be useful in the
following, we are going to sketch it. In this section we consider
that all the maps between smooth manifolds are smooth; for
equivalence between two principal $\H$-bundles over $\Sigma$ we
mean a bundle morphism which is a diffeomorphism on the total
space and induces the identity on $\Sigma$ and on $\H$.

Let $P=P(\Sigma,\H)$ be a flat principal bundle over $\Sigma$ with
fibre $\H$, {\it i.e.} a bundle that admits a flat connection. It
is always possible to choose an open covering $\{U_\alpha\}$ for
$\Sigma$ and the trivialization for $P$ such that the transition
functions $h_{\alpha\beta}$ are constant. In fact let $\rho\in
Hom(\pi_1(\Sigma), \H)$ be the map that associates to each cycle
the holonomy around it (defined up to conjugation). It can be
shown that $P$ is equivalent to $\tilde{\Sigma}\times_\rho\H$,
where $\tilde{\Sigma}$ is the universal covering of $\Sigma$, and
that $\tilde{\Sigma}\times_\rho\H$ admits a canonical locally
constant trivialization, see \cite{DeBa} for details. Let
$h_{\alpha\beta}$ be the transition functions in this
trivialization of $P$; we can describe all flat connections in $P$
as $J=\{-d\psi_\alpha \psi_\alpha^{-1}\}$, with
$\psi_\alpha:U_\alpha\rightarrow \H$. The compatibility conditions
imply that \beq
 h_{\alpha\beta}d\psi_\beta
\psi_{\beta}^{-1} h_{\beta\alpha}= d\psi_\alpha\psi_\alpha^{-1}
\eeq{tansfunc}
 on $U_\alpha \cap  U_\beta$. We can associate to
the collection of $\{\psi_\alpha\}$ an equivalent bundle
$\hat{P}(\Sigma,\H)$ with transition functions
$h^J_{\alpha\beta}= \psi_{\alpha}^{-1}h_{\alpha\beta}\psi_\beta$
on $U_\alpha \cap U_\beta$ which due to (\ref{tansfunc})
satisfy
\beq
 \psi_\alpha  d h^J_{\alpha\beta} \psi_\beta^{-1} =
-d\psi_\alpha \psi_\alpha^{-1} h_{\alpha\beta} +
 h_{\alpha\beta} d\psi_\beta \psi_\beta^{-1} = 0
\eeq{consthJJ}
 and therefore $h^J_{\alpha\beta}$ are constants and
define a new locally constant trivialization of $P$. Let us
identify $g^{-1}_{\alpha} h^J_{\alpha\beta}g_\beta$ and
$h^J_{\alpha\beta}$ for all constant $g_\alpha\in \H$ and denote
the corresponding equivalence class $[h^J]$. We denote by ${\cal
F}(\Sigma, \H, P)$ the moduli space of gauge inequivalent flat
connections on $P$. Let us denote with $[J]$ the class of $J$ in
${\cal F}(\Sigma, \H,P)$. The gauge transformations are defined as
automorphisms of $P$, i.e. $\gamma\in{\rm Aut}(P)$ is defined by
$\gamma=\{\gamma_\alpha\,|\, \gamma_\alpha: U_\alpha\rightarrow
\H\}$ such that $\gamma_\alpha
h_{\alpha\beta}=h_{\alpha\beta}\gamma_\beta$ on $U_\alpha \cap
U_\beta$. Then
$\gamma(J)=\{-d\psi_\alpha^\gamma(\psi_\alpha^\gamma)^{-1}\}$ with
$\psi_\alpha^\gamma=\gamma_\alpha\psi_\alpha$. Since
$h^{\gamma(J)}_{\alpha\beta}=h^{J}_{\alpha\beta}$, we get a well
defined map that sends $[J]$ in $[h^J]$.

Let us suppose that $\hat{h}$ are locally constant transition
functions that define a bundle $\hat{P}(\Sigma,
\H)$ equivalent to $P$, then it exists
$\{\psi_{\alpha}:U_\alpha\rightarrow \H \}$ such that
$\hat{h}_{\alpha\beta}=\psi_{\alpha}^{-1}h_{\alpha\beta}\psi_\beta$.
It is then clear that
$J^{\hat{h}}=\{-d\psi_\alpha\psi_\alpha^{-1}\}$ is a flat
connection in $P$ and that the map that sends $\hat{h}$ in
$J^{\hat{h}}$ depends only on $[\hat{h}]$. Every class of flat
connections $[J]$ in $P$ can then be represented by the class of
locally constant transition functions $[h^J]$. In other words
 the space ${\cal F}(\Sigma, \H, P)$ can be equivalently defined as follows
\beq
{\cal F}(\Sigma,\H, P)=
\{\hat{h}_{\alpha\beta}:U_\alpha\ \cap U_\beta\rightarrow\H
\,|\,\hat{h}_{\alpha\beta}\hat{h}_{\beta\gamma}=\hat{h}_{\alpha\gamma},\,
d\hat{h}_{\alpha\beta}=0\}/
\sim
\eeq{definFwithbbound}
where we consider only those constant $\hat{h}_{\alpha\beta}$ which define the bundle $\hat{P}$
  equivalent to $P$ while
 the equivalence $\sim$ is defined as
 $\hat{h}_{\alpha\beta}\sim s_\alpha^{-1}\hat{h}_{\alpha\beta}s_\beta$,
 for a constant $s_\alpha\in \H$.
 We refer to
\cite{Ivanova:1999iq} for the interpretation of $[h]$ in the
\v{C}ech cohomology of the sheaf of locally constant sections on
${\rm Ad} P$, where ${\rm Ad}P=P\times_G G$. Finally we can define the space
\begin{equation}
\label{inequivbundles}
{\cal F}(\Sigma,\H)=\bigcup_{[P]} {\cal F}(\Sigma,\H, P)
\end{equation}
where $P=P(\Sigma,\H)$ is a representative in the class $[P]$ of
equivalent bundles and the union runs over all these classes.

 Now let us try to apply the same logic to the case of Riemann surfaces
 with boundary. For the sake of clarity let us assume that
 the boundary $\d\Sigma$ has a single component which is homeomorphic
 to $S^1$. Let ${\cal K}$ be a subgroup of $\H$, $P(\d\Sigma,{\cal K})$ be a ${\cal K}$-bundle
over $\d\Sigma$ and $P(\Sigma,\H)$ be an ${\cal H}$-bundle over $\Sigma$. We require to exist
a bundle morphism
\beq
 P(\d\Sigma, {\cal K})\,\,\,\rightarrow\,\,\,P(\Sigma, {\cal H}),
\eeq{defprbboun}
 which implies the injections $\d\Sigma \hookrightarrow \Sigma$ and ${\cal K}
 \hookrightarrow {\cal H}$.  Let us choose a good open covering $\{ U_{\alpha}\}$
 of $\Sigma$ such that  $\{V_\alpha = U_\alpha \cap \d\Sigma\ \neq \emptyset\}$
 is a covering of $\d\Sigma$. We assume that there exists such trivialization for
 the flat bundle $P(\Sigma, {\cal H})$ such that the bundle map (\ref{defprbboun})
 is realized locally as the injection
\beq
 V_\alpha \times {\cal K} \,\,\,\hookrightarrow\,\,\, U_\alpha \times {\cal H}.
\eeq{localbmap}
This means that if $h_{\alpha\beta}$ are the transition functions for $P$ in this trivialization
then $h_{\alpha\beta}|_{\d\Sigma}\in{\cal K}$ define the transition functions
for $P(\d\Sigma, {\cal K})$. We denote all these data as $P(\Sigma,\H,{\cal K})$.

There is a natural notion of equivalence of bundles with such
boundary conditions. We say that $P(\Sigma,\H,{\cal K})$ is
equivalent to $\tilde{P}(\Sigma,\H,{\cal K})$ if there exist maps
$\xi_\alpha : U_\alpha \rightarrow {\cal H}$ and
 $\xi_{\alpha}|_{V_\alpha} \in {\cal K}$ such that
$\tilde{h}_{\alpha\beta} = \xi^{-1}_\alpha h_{\alpha\beta} \xi_\beta$, where $h_{\alpha\beta}$
(resp.$\tilde{h}_{\alpha\beta}$) are the transition functions for $P$ (resp.$\tilde{P}$).
This is equivalent to say that the following diagram is commutative
\beq
\begin{array}{lll}
  P(\d\Sigma, {\cal K}) &  \rightarrow  &  P(\Sigma, {\cal H}) \\
 \,\,\,\,\,\,\,\,\,\,   \updownarrow &    & \,\,\,\,\,\,\,\,\,\,\,\updownarrow\  \\
  \tilde{P}(\d\Sigma, {\cal K}) &  \rightarrow  & \tilde{P}(\Sigma, {\cal H}) \;,
\end{array}
\eeq{equivbundle}
where the vertical arrows denote the standard equivalence of bundles defined by $\xi_\alpha$ and
$\xi_\alpha|_{\d\Sigma}$.

We are going to define flat connections in $P(\Sigma,\H,{\cal K})$. Namely we consider
flat connections $J$ in $P(\Sigma, {\cal H})$ whose restriction on the boundary
$\d\Sigma$ reduces to connections over $P(\d\Sigma, {\cal K})$.
A gauge transformation is an automorphism of $P(\Sigma,\H,{\cal K})$,
{\it i.e.} it is defined as
$\gamma = \{ \gamma_\alpha| \gamma_\alpha: U_\alpha \rightarrow {\cal H},\,\,\gamma_\alpha|_{V_\alpha}
 \in {\cal K}\}$ such that $\gamma_\alpha h_{\alpha\beta} = h_{\alpha\beta} \gamma_\beta$ on
 $U_\alpha \cap U_\beta$.
Now it is a straightforward exercise to generalize
the description of flat connections with prescribed boundary conditions in terms
of the flat structures described in (\ref{defprbboun}). In the fixed trivialization of (\ref{defprbboun})
 we can describe all flat connections as $J= \{-d\psi_\alpha \psi_\alpha^{-1}\}$, with $\psi_\alpha: U_\alpha
 \rightarrow {\cal H}$ and $\psi_\alpha|_{V_\alpha} \in {\cal K}$. We can construct an
equivalent bundle $ \hat{P}(\Sigma,\H,{\cal K})$ with
 the constant transition functions $h^J_{\alpha\beta} = \psi_{\alpha}^{-1} h_{\alpha\beta} \psi_\beta$.
  These transition functions define the following space
\beq
{\cal F}(\Sigma,\H,{\cal K}, P)=
\{\hat{h}_{\alpha\beta}:U_\alpha\cap U_\beta\rightarrow\H
\,|\,\hat{h}_{\alpha\beta}\hat{h}_{\beta\gamma}=\hat{h}_{\alpha\gamma},\,
d\hat{h}_{\alpha\beta}=0,\,\hat{h}_{\alpha\beta}|_{\d\Sigma}\in{\cal K}, \}/
\sim
\eeq{defspacefjkl}
where we consider only those constant $\hat{h}_{\alpha\beta}$ which define bundles $\hat{P}(\Sigma,\H,{\cal K})$
equivalent to $P(\Sigma,\H,{\cal K})$ according to (\ref{equivbundle}), while
the equivalence $\sim$ is defined as
 $\hat{h}_{\alpha\beta}\sim s_\alpha^{-1}\hat{h}_{\alpha\beta}s_\beta$,
for constant $s_\alpha\in \H$ such that  $s_\alpha|_{V_\alpha} \in {\cal K}$.
In analogy with the previous discussion for the case without boundary one can
show that two gauge equivalent flat connections on (\ref{defprbboun}) define the same element
in ${\cal F}(\Sigma,\H,{\cal K}, P)$.
Also we can go in the opposite direction: an element $\hat{h}_{\alpha\beta}$ of
${\cal F}(\Sigma,\H,{\cal K}, P)$ defines a flat bundle $\hat{P}(\Sigma,\H,{\cal K})$
which is equivalent to $P(\Sigma,\H,{\cal K})$, {\it i.e.} there exist
$\{\psi_\alpha : U_\alpha \rightarrow {\cal H},\,\,
 \psi_\alpha|_{V_\alpha} \in {\cal K}\}$ such that $\hat{h}_{\alpha\beta} = \psi_\alpha^{-1} h_{\alpha\beta}
 \psi_\beta$. Then one can construct the flat connection $J=\{-d\psi_\alpha \psi_\alpha^{-1}\}$.
Thus ${\cal F}(\Sigma,\H,{\cal K}, P)$ is
the space of gauge inequivalent  flat connections on $P(\Sigma,\H,{\cal K})$. Finally  we can define the space
\begin{equation}\label{defspaceall}
{\cal F}(\Sigma,\H,{\cal K}) = \bigcup_{[P]}{\cal F}(\Sigma,\H,{\cal K}, P)
\end{equation}
where $P=P(\Sigma,\H,{\cal K})$ is a representative of the class
$[P]$ of equivalent bundles $P(\Sigma,\H,{\cal K})$ and the union
is over all these classes.

In the next two subsections we will use this description of flat connections.
This description is an useful tool to deal with solutions that correspond to
topologically non trivial bundles. In fact, even
 assuming that the group  $\G^*$ is connected and
simply connected we still have to consider stability subgroups of symplectic leaves
 which can be not simply connected and therefore also admit non trivial principal bundles.

\subsection{Riemann surfaces without boundaries}
\label{no_boundary}

In this subsection we consider a Riemann surface $\Sigma$ without
boundary, i.e. $\d\Sigma = \emptyset$. We consider the case when
$\G$ and $\G^*$ are connected and simply connected groups. In this
situation any principal bundle with base $\Sigma$ and fiber $\G^*$
is equivalent to the trivial one, $\Sigma\times\G^*$. The group
of gauge transformations is
$G(\Sigma,\G^*)=\{\gamma:\Sigma\rightarrow\G^*\}$. Let us consider
$\Sigma\times\G$ as the fibre bundle associated to the dressing
action of $\G^*$ on $\G$. In the linear case it is a vector
bundle, in the general case it is just a fibre bundle. Thus the
relevant fields $(X, J)$ can then be described as a connection $J$
on the principal bundle $\Sigma \times \G^*$ and as a section
$\hat{X}(\xi)=(\xi,X(\xi))$ of the associated fiber bundle $\Sigma\times
\G$. The infinitesimal transformations defined in
(\ref{newgautran1}) and (\ref{newgautran}) are integrated to the
action of $G(\Sigma,\G^*)$  on the
 solutions  $(X, J)$.   We then define the main object of our study:
\begin{equation}\label{moduli_space}
M(\Sigma,\G)= \frac{\{{\rm Solutions ~of~
(\ref{neweqofmot})}\}}{G(\Sigma,\G^*)} \;.
\end{equation}
If $(X,J)$ is a solution of equations of (\ref{neweqofmot}) let us
indicate with $[(X,J)]$ the corresponding element in
$M(\Sigma,\G)$.

It is easy to find the local solution of (\ref{neweqofmot}). Let
us choose some open covering $\{ U_\alpha \}$ of $\Sigma$. Then
since $J$ is flat, we can always find
$\psi_\alpha:U_\alpha\rightarrow\G^*$ such that
$J=\{-d\psi_\alpha\psi_\alpha^{-1}\}$ and moreover there exist
$x_0^\alpha\in\G$ such that $X(\xi)=\{\psi_\alpha(\xi)(x_0^\alpha)\}$
on $U_\alpha$, where $\psi_\alpha$ acts on $x_0^\alpha$ by means
of the dressing transformation. Since we deal with the trivial
bundle there are the following gluing conditions
\beq
d\psi_\alpha\psi_\alpha^{-1}=d\psi_\beta\psi_\beta^{-1},\,\,\,\,\,\,\,\,\,
 \psi_\alpha(\xi)(x_0^\alpha) = \psi_\beta (\xi)(x_0^\beta),\,\,\,\,\,\,\,\,\,
{\rm on}\,\, U_\alpha \cap U_\beta .
\eeq{compatib}
Since the
dressing action preserves the symplectic leaves of $\G$, we can
conclude that $x_0^\alpha$ and $X(\xi)$ stay inside the same
symplectic leaf $\Sy$ (for the general statement see Appendix).
 Let us fix a point $x_0\in\Sy$. Since the
dressing action is transitive on $\Sy$ we can always find
$s^\alpha\in\G^*$ such that $x_0^\alpha=s^\alpha(x_0)$. If we
define $\phi_\alpha=\psi_\alpha\circ s^\alpha$ then
$X(\xi)=\{\phi_\alpha(x_0)\}$ and
$J=\{-d\phi_\alpha\phi_\alpha^{-1}\}$, {\it i.e.} due to
(\ref{compatib}) in the class $[h^J]$ describing $J$ we can choose
a representative
$h^{X,J}_{\alpha\beta}=\phi_\alpha^{-1}\phi_\beta\in \H_{(\Sy,
x_0)}$, the stabilizer subgroup of $x_0$. If we change $s^\alpha$
in $s^\alpha x^\alpha$ with $x^\alpha\in \H_{(\Sy, x_0)}$ we get
$x_\alpha^{-1}h^J_{\alpha\beta}x_\beta$. So we have defined a
mapping that sends $[(X,J)]$ to $[h^{X,J}]\in{\cal F}(\Sigma,\H_{(\Sy,x_0)})$:
this statement is
equivalent to saying that $J$ reduces to an $\H_{(\Sy,
x_0)}$-connection. Next we show that the map $[X,J]\rightarrow
[h^{X,J}] $ is injective. In fact suppose that
$(X=\psi_\alpha(x_0),J=-d\psi_\alpha\psi_\alpha^{-1})$ and
$(Y=\phi_\alpha(x_0),J'=-d\phi_\alpha\phi_\alpha^{-1})$ are mapped
to the same flat connection $[h_{\alpha\beta}]$. This means that
there exists $x_\alpha\in \H_{(\Sy, x_0)}$ such that
$\psi_\alpha^{-1}\psi_\beta =
x_\alpha^{-1}\phi_{\alpha}^{-1}\phi_\beta x_\beta$. It is then
easy to verify that $X=\gamma(Y)$ and $J=\gamma(J')$ with
$\gamma=\psi_\alpha x_\alpha^{-1}\phi_\alpha^{-1}=\psi_\beta
x_\beta^{-1}\phi_\beta^{-1}$.

Let $[\hat{h}_{\alpha\beta}]$ describe a flat $\H_{(\Sy,
x_0)}$-connection living in the bundle defined by constant
$h_{\alpha\beta}$; this means that there exist
$\phi_\alpha:U_\alpha\rightarrow \H_{(\Sy,x_0)}$ such that
$\hat{h}_{\alpha\beta}=\phi_\alpha^{-1}h_{\alpha\beta}\phi_\beta$.
Obviously $[\hat{h}_{\alpha\beta}]$ defines also a flat
$\G^*$-connection, that, being $\G^*$ simply connected, lives in a
bundle equivalent to $\Sigma\times\G^*$. Then there exists
$\psi_\alpha:U_\alpha\rightarrow \G^*$ such that
$h_{\alpha\beta}=\psi_\alpha^{-1}\psi_\beta$ and
$J=\{-d(\psi_\alpha\phi_\alpha)(\psi_\alpha\phi_\alpha)^{-1}\}$
and $X(u)=\psi_\alpha\phi_\alpha(\xi)(x_0)$ is a solution of
(\ref{neweqofmot}).

The space of inequivalent solutions $(X,J)$, such that $X$ lives
in $\Sy$, is in one to one correspondence with the moduli space of
flat $\H_{(\Sy, x_0)}$ connections on $\Sigma$ (including the
reducible ones!). If $\H_{(\Sy, x_0)}$ is not a simply connected
subgroup of $\G^*$ then we will have to take into account all the
inequivalent bundles $P(\Sigma, \H_{(\Sy,x_0)})$ and we will get
the whole space \beq {\cal F}(\Sigma, \H_{(\Sy, x_0)})=
\bigcup\limits_{[P]}{\cal F}(\Sigma, \H_{(\Sy, x_0)},P) \;.
\eeq{spacsingle} In the case when $\H_{(\Sy, x_0)}$ is simply
connected then every principal bundle is equivalent to the trivial
one and the space (\ref{spacsingle}) is just the space of the flat
connections for $\Sigma \times \H_{(\Sy, x_0)}$. In general one
can verify that it is possible to find a gauge such that $X=x_0$
and $J$ is an $\H_{(\Sy, x_0)}$-connection form if and only if the
$\H_{(\Sy, x_0)}$-bundle defined by $h^{X,J}_{\alpha\beta}$ is
trivial. In fact, with the same notations than before for $(X,J)$,
if there exists $f_\alpha:U_\alpha\rightarrow\H_{(\Sy, x_0)}$ such
that $h^{X,J}_{\alpha\beta}=f_\alpha^{-1}f_\beta$ then
$\gamma=f_\alpha\phi_\alpha^{-1}=f_\beta\phi_\beta^{-1}$ extends
to all $\Sigma$ and defines the desired gauge transformation.

Since dressing transformations preserve symplectic leaves, gauge
transformations cannot mix solutions living in different leaves.
Therefore one can conclude that the whole moduli space is the
union over symplectic leaves $\Sy$ of the spaces described in
(\ref{spacsingle}). All the previous discussion can be summarized
in the following theorem:

\begin{proposition}\label{p:noboundary}
Let $L_\G$ be the space of symplectic leaves of $\G$ and $p:
M(\Sigma,\G)\rightarrow L_\G$ be the map that associates to $[X,J]$
the symplectic leaf where $X$ lives. Then, for each ${\cal S}\in
L_\G$, fix $x_0\in{\cal S}$, we have that
$$
p^{-1}({\cal S}) = {\cal F}(\Sigma, \H_{(\Sy, x_0)}) \;,
$$
where  $\H_{(\Sy, x_0)}\subset\G^*$ is the stability subgroup of $x_0$.
\end{proposition}

This shows that the space $M(\Sigma,\G)$ is a topological space
with the topology induced from the space of symplectic leaves
$L_\G$ which is equipped with the quotient topology. However we
should admit that this topology on $M(\Sigma,\G)$ is too rough. We feel that the
present level of discussion is too
general and extra conditions should be introduced in order
  to define a finer topology on $M(\Sigma,\G)$.

\begin{remark}\rm The analysis of equations of motion
can be done in an alternative and very geometrical way. The first
equation of (\ref{neweqofmot}) means that $X$ defines a parallel
section $\tilde{X}(u)=(u,X(u))$ with respect to the connection
$J$. In fact let $\omega_J(u,\gamma)\in(T^*_u\Sigma\oplus
T_\gamma^* \G^*)\otimes\g^*$ be the connection form defined by $J$
and let $H_{(u,\gamma)}=\{v\in T_u\Sigma\oplus T_\gamma\G^*,
\langle \omega_J,v\rangle = 0\}$ be the horizontal space. Let
$g\in\G$ define the map
$g:\Sigma\times\G^*\rightarrow\Sigma\times\G$,
$g(u,\gamma)=(u,\gamma(g))$ and let
$\tilde{H}_{u,g}=g_*(H_{(u,e)})$ define the horizontal space in
the associated bundle $\Sigma\times\G$. It is straightforward to
verify that the the first of (\ref{neweqofmot}) is equivalent to
$\tilde{X}_*(T\Sigma)\subset\tilde{H}$. Moreover the first of
(\ref{neweqofmot}) implies also that $X(\Sigma)$ is contained in a
single symplectic leaf $\Sy=\G^*/\H_{(\Sy, x_0)}$ (see Appendix), for some
arbitrary $x_0\in\Sy$. Then $\tilde{X}$ defines a
section in the associated fibre bundle $\Sigma\times\G^*/\H_{(\Sy,
x_0)}$. If now we apply Prop.7.4 of \cite{KN} we can conclude that
$\tilde{X}$ is parallel if and only if $J$ reduces to a connection
in the $\H_{(\Sy, x_0)}$-bundle $P_X(\Sigma,\H_{(\Sy,
x_0)})=\{(u,\gamma)\in\Sigma\times\G^*\,|\, X(u)=\gamma(x_0)\}$.
Remark that $P_X$ is the pullback by $X$ of the homogeneous bundle
$\G^*\rightarrow\G^*/\H_{(\Sy, x_0)}$.
\end{remark}

\medskip
\begin{example}{\rm
Let us consider the moduli space over the two dimensional
sphere $S^2$. In this case all flat connections are equivalent to
the trivial one whatever is the group $\H_{(\Sy,x_0)}$ and therefore the moduli space of solutions
coincides with the space of the symplectic leaves for $\G$, {\it i.e.} $M(S^2, \G)=L_\G$.
In general this space can be even non-Hausdorf.} \end{example}

\subsection{Riemann surfaces with boundaries}\label{mboundary}

Now we turn to the description of the moduli space of
solutions over a Riemann surface with boundary. For the sake of
clarity we assume that the boundary has a single component, which we
denote $\d \Sigma$. However the generalization for the case with
more than one component is straightforward.

In the previous subsection we have solved problem in the bulk and
now we have to incorporate the boundary conditions. Let us remind the
discussion of boundary conditions from Section \ref{s:group}. We have argued that the
appropriate conditions are given by a coisotropic submanifold
$\D$ which is invariant under coisotropic subgroups $\H\subset\G$ and
$\H^\perp\subset\G^*$. For the fields the following conditions are imposed
\beq
X :\partial\Sigma\rightarrow \D,\,\,\,\,\,\,\,\,
J|_{\partial\Sigma}\in\Omega^1(\partial\Sigma)\otimes \h^\perp
\eeq{definfbjc}
where $\h^\perp$ is the  Lie algebra of the
subgroup $\H^\perp\subset\G^*$. On the boundary the infinitesimal
gauge transformations defined by $\beta=\beta_A T^A$ are
  living in $\h^\perp$. Therefore we can interpret
$J|_{\d\Sigma}$ as a connection for the trivial $\H^\perp$
principal bundle over $\d\Sigma \sim S^1$.

We assume that $\G^*$ is simply connected and
$\H^\perp$ is connected.  We consider the trivial
bundles with the natural injection
\beq
 \d\Sigma\times {\cal H}^\perp\,\,\,\hookrightarrow\,\,\,\Sigma \times {\cal G}^*
\eeq{trivboundb}
Accordingly the gauge group is $G(\Sigma,\G^*,
\H^\perp)=\{\gamma:\Sigma\rightarrow\G^*\,,\,
\gamma|_{\partial\Sigma}\in \H^\perp\}$. Following the discussion
 in subsection \ref{modfl}  $J$ is interpreted as a connection on
(\ref{trivboundb}) and  $X$ is a section of
the associated bundle $\Sigma\times\G$ that restricts on the
boundary to a section of $\partial\Sigma\times \D$ (which can be seen as
associated to $\d\Sigma\times {\cal H}^\perp$).  We define the
moduli space of the solutions
\begin{equation}
M(\Sigma,\G,\H^\perp,\D) = \frac {\{{\rm Solutions~
of~(\ref{neweqofmot})},\, X: \partial\Sigma \rightarrow \D,
J|_{\partial\Sigma}\in\Omega^1(\partial\Sigma)\otimes\h^\perp\}}{G(\Sigma,\G^*, \H^\perp)}\,,
\end{equation}
which we describe below.

 In the bulk the analysis of (\ref{neweqofmot}) is exactly the same
as in previous subsection: the $X$ field lives in a symplectic leaf $\Sy$
and the connection $J$ reduces to the stability subgroup
$\H_{(\Sy, x_0)}$, for some $x_0\in\Sy$. The boundary conditions
force $X|_{\d\Sigma}$ to live in $\Sy\cap\D$. Therefore we have
to consider only those symplectic leaves such that $\Sy \cap \D
\neq \emptyset$. Moreover the first equation of (\ref{neweqofmot})
has a well defined restriction to the boundary $\d\Sigma$,
\beq
(\d_\tau X^\mu - s^{A\mu} J_{\tau A})|_{\d\Sigma} =0,
\eeq{alongthebc}
 where $\tau$ parametrizes the boundary. Since
$J_{\tau}|_{\d\Sigma}$ lives in $\h^\perp$ then $X|_{\d\Sigma}$
lies entirely in a single orbit ${\cal O} (\Sy,
\H^\perp)\subset\Sy$ of $\H^\perp$ on $\D$. Since gauge
transformations are restricted on the boundary to $\H^\perp$, each
point in the moduli space of solutions identifies in this way an
$\H^\perp$-orbit in $\D$. Let us choose $x_0\in{\cal O} (\Sy,
\H^\perp)\subset\Sy$; then ${\cal O} (\Sy,
\H^\perp)=\H^\perp/(\H^\perp\cap\H_{(\Sy, x_0)})$. With the
same mechanism as in the bulk, on the boundary $J|_{\d\Sigma}$ reduces to an
$\H^\perp\cap\H_{(\Sy, x_0)}$-connection.

Let us spell out more details of this construction.
We introduce the local
trivialization of $J=\{-d\psi_\alpha \psi_\alpha^{-1}\}$, where
$\psi_\alpha:U_\alpha\rightarrow\G^*$. As a consequence of
boundary conditions, $\psi_\alpha$ should be chosen such that
$\psi_\alpha|_{\partial\Sigma}\in \H^\perp$. Then we have that
$X(u)=\psi_\alpha(x_0^\alpha)$, so that $x_0^\alpha$ stay in
${\cal O} (\Sy, \H^\perp)$ for $\alpha$ such that $V_\alpha=U_\alpha\cap\d\Sigma\not=\emptyset$.
Let us define
$s^\alpha_0\in\G^*$ such that $x_0^\alpha=s_0^\alpha(x_0)$. By
construction, we can choose $s_0^\alpha\in \H^\perp$ for all
$\alpha$ such that $V_\alpha\not=\emptyset$. Let us define $\phi_\alpha=\psi_\alpha
s^\alpha_0$: it is clear that $\phi_\alpha|_{\partial\Sigma}\in
\H^\perp$ for all $\alpha\in{\cal I}$. We have that
$h_{\alpha\beta}^{X,J}=\phi_\alpha^{-1}\phi_\beta\in \H_{(\Sy,
x_0)}$, and $h_{\alpha\beta}^{X,J}|_{\d\Sigma}\in \H_{(\Sy,
x_0)}\cap \H^\perp$. We get a map that sends $[(X,J)]$ to
$[h^{X,J}]\in{\cal F}(\Sigma,\H_{(\Sy, x_0)},\H_{(\Sy,
x_0)}\cap \H^\perp)$, where this space has been defined in (\ref{defspaceall}). It is easy to verify
that this map is injective.

Let us discuss now the inverse map. Let $[h_{\alpha\beta}]\in{\cal
F}(\Sigma,\H_{(\Sy, x_0)},\H_{(\Sy, x_0)}\cap \H^\perp, P)$,
 where $P$ denotes a bundle defined as in (\ref{defprbboun}). Since
$\G^*$ is simply connected, the $\G^*$ bundle on $\Sigma$ defined
by $h_{\alpha\beta}$ is equivalent to the trivial one, {\it i.e.}
there exists $\psi_\alpha:U_\alpha\rightarrow\G^*$ such that
$h_{\alpha\beta}=\psi_\alpha^{-1}\psi_\beta$. Analogously since
$\H^\perp$ is connected the $\H^\perp$-bundle on the boundary is
trivial so that there exists
$\phi_\alpha:V_\alpha\rightarrow\H^\perp$ such that
$h_{\alpha\beta}|_{\d\Sigma}=\phi_\alpha^{-1}\phi_\beta$. The map
$\gamma_\alpha=\psi_\alpha|_{\d\Sigma}\phi_\alpha^{-1}:
V_\alpha\rightarrow\G^*$ coincides on $V_\alpha\cap V_\beta$ so
that it can be extended to $\gamma:S^1\rightarrow\G^*$. This means
that $J^h=-d\psi_\alpha\psi_\alpha^{-1}$ is a $\G^*$-connection on
$\Sigma\times\G^*$ whose restriction to the boundary is gauge
equivalent to $A=-d\phi_\alpha\phi^{-1}_\alpha$, {\it i.e.}
$J^h|_{\d\Sigma}=\gamma(A)$. Since $\G^*$ is simply connected
every $\gamma:S^1\rightarrow\G^*$ can be extended to
$\tilde{\gamma}:\Sigma\rightarrow\G^*$, such that
$\tilde{\gamma}|_{\d\Sigma}=\gamma$. Then
$J=\tilde{\gamma}^{-1}(J^h)=-d(\tilde{\gamma}^{-1}\psi_\alpha)
(\tilde{\gamma}^{-1}\psi_\alpha)^{-1}$,
$X=\tilde{\gamma}^{-1}\psi_\alpha(x_0)$ is a solution of
(\ref{neweqofmot}) satisfying boundary conditions.

We have shown that the space of inequivalent solutions satisfying
the boundary conditions (\ref{definfbjc}) for a fixed
orbit ${\cal O}(\Sy,\H^\perp)\subset \D$, is in one to one
correspondence with the space of $\H_{(\Sy,x_0)}$ flat connections satisfying
$\H_{(\Sy, x_0)}\cap \H^\perp$ boundary conditions, {\it i.e.}
\beq
{\cal F}(\Sigma,\H_{(\Sy, x_0)},\H_{(\Sy,
x_0)}\cap \H^\perp)=
 \bigcup\limits_{[P]} {\cal F}(\Sigma,\H_{(\Sy, x_0)},\H_{(\Sy,
x_0)}\cap \H^\perp, P),
\eeq{resultbcsuno}
where again $[P]$ runs over the inequivalent bundles $P(\Sigma,\H_{(\Sy, x_0)},\H_{(\Sy,
x_0)}\cap \H^\perp)$. Since dressing transformations preserve
symplectic leaves and $\H^\perp$-orbits in $\D$, the whole moduli
space of solutions will be given by the union over the different
$\H^\perp$-orbits in $\D$. Let us summarize this fact in the
following proposition.

\begin{proposition}\label{p:boundary}
Let $L_\D$ be the space of $\H^\perp$-orbits inside $\D$ and
$p:M(\Sigma,\G,\H^\perp,\D)\rightarrow L_\D$ be the map that
associates to $[X,J]$ the orbit where $X|_{S^1}$ lives. Then for
each ${\cal O}(\Sy,\H^\perp)\in L_\D$ fix $x_0\in{\cal
O}(\Sy,\H^\perp)$. We have that
$$
p^{-1}({\cal O}(\Sy,\H^\perp)) = {\cal F}(\Sigma,\H_{(\Sy,
x_0)},\H_{(\Sy, x_0)}\cap \H^\perp)\;,
$$
 with $\H_{(\Sy, x_0)}$ being the stability subgroup of $x_0$.
\end{proposition}

\begin{example}\label{disk}{\rm  Let us consider the model defined on the disk $D$.
For the disk $D$ all flat connections are gauge equivalent to the trivial one and therefore
the moduli space of solutions coincides with the space of $\H^\perp$ orbit of $D$, {\it i.e.}
$M(D, \G, \H^\perp, \D)=L_{\cal D}$.}
\end{example}

\section{BF-theory}
\label{s:bf}

In this section we would like to describe a concrete example and
illustrate the general results for the moduli spaces given in
Section \ref{s:moduli}. We consider here the case of the linear
Poisson structure which is usually called BF-theory.  The
BF-theory is known for 15 years \cite{Horowitz:ng, Blau:1989bq}
and is relatively well studied. At the same time it is one of the
simplest non-trivial examples of the Poisson sigma model.

First we briefly remind the description of the BF-model to show that it is really a Poisson-Sigma model
on a Poisson-Lie group. The model is defined on a vector space $\G$, that we consider as
the abelian group of translations. We introduce on it the linear Poisson structure $ \alpha^{BC}=
\tilde{f}^{BC}_{\,\,\,\,\,\,\,\,\,A}X^A$ where $ \tilde{f}^{BC}_{\,\,\,\,\,\,\,\,\,A}$ are the structure constants
for some group $\G^*$ such that $\dim\G^*=\dim\G$. The group $\G^*$  acts on $\G$ by the coadjoint action
and thus $\G$ can be identified with the dual space of the Lie algebra of $\G^*$.
The space $\G$ with the Poisson structure $\alpha^{AB}$ is a Poisson-Lie group and it is dual to $\G^*$ equipped
with the trivial Poisson-Lie structure.  The dressing vector fields are the fundamental vector fields of the
coadjoint action of $\G^*$ on $\G$ so that they are complete by construction.
The model is given by the following action
\beq
 S = \int\limits_{\Sigma}
  \,\,X^A\left (  d \eta_{A}
 + \frac{1}{2} \tilde{f}^{BC}_{\,\,\,\,\,\,\,\,\,A} \eta_B \wedge \eta_{C} \right ) ,
\eeq{BFtheacdef11}
where $\eta$ is a connection on the trivial bundle $\Sigma\times\G^*$ and $X: \Sigma \rightarrow \G$ is
interpreted as a section of the vector bundle $\Sigma\times\G$  associated to the coadjoint action. As it was
  pointed out
at the end of Section \ref{s:general} (\ref{BFtheacdef11}) differs
from the action of the Poisson Sigma model by a boundary term.
However this fact does not affect the boundary conditions. Under
the gauge symmetries (\ref{gaugetransf}) $X$ transforms as a
section of $\Sigma\times \G$
  and $\eta$ as a $\G^*$ connection.
The equations of motions for (\ref{BFtheacdef11}) imply that $\eta$ is a flat $\G^*$ connection and
$X$ is a covariantly constant section of the vector bundle. In particular $\eta$ needs not to be redefined, $\it i.e.$
$\eta=J$ in (\ref{newvar}).

The symplectic leaves on $\G$ are then the coadjoint orbits of
$\G^*$ and one can in principle apply the results of orbit method
to get informations about the moduli space of solutions
$M(\Sigma,\G)$ (see \cite{KB}). Let us assume for instance that
$\G^*$ is a compact (connected and simply connected) group. The
stability subgroup of each coadjoint orbit ${\cal S}$ is a
subgroup $\H_{(\Sy, x_0)}$ which is contained in a finite number
of conjugacy classes of subgroups of $\G^*$. In particular the
stability subgroup for the orbits of maximal dimension can be
chosen as a maximal abelian connected subgroup $T$ of $\G^*$ and
for the generic orbit it can be chosen such that
$T\subset\H_{(\Sy, x_0)}\subset\G^*$. There is a finite number of
typical fibers of the projection map described in Proposition
\ref{p:noboundary} and the fibers over the orbits of maximal
dimension are all isomorphic.

\begin{example}\label{example_no_boundary}{\rm
Let us be more concrete and consider the very simple case when $\G^*=SU(2)$ and $\Sigma$ is closed.
We refer to the notations of Proposition \ref{p:noboundary}.
Here $\G={\bf R}^3$ can be seen as the additive group of translations and the dressing transformations
are just rotations on ${\bf R}^3 $.
A coadjoint orbit is identified by the equation $X^A \eta_{AB} X^B = \rho$, where
$\eta_{AB}$ is the Cartan-Killing metric and $\rho\geq 0$. The space
  $L_\G$ of symplectic leaves is then ${\mathbf R}^+$. The case $\rho=0$ is a degenerate point
and its stability subgroup is the whole $SU(2)$: so
$p^{-1}(0)={\cal F}(\Sigma,SU(2))$. If $\rho > 0$ the orbit is a
sphere and the stability subgroup is $U(1)$, so that
$p^{-1}(\rho)={\cal F}(\Sigma,U(1))$. Since $U(1)$ is not simply
connected group, we have to take into account also the
contribution of topologically non trivial $U(1)$ bundles. The
solutions corresponding to the trivial $U(1)$ bundle can be always
put in the form where $X=x_0$ and $J$ is a flat $U(1)$ connection
form. However it should be stressed that there exist also the
solutions associated to the non trivial $U(1)$ bundles and these
solutions cannot be written in this simple form.  As an example of
discussion of the flat connections on the two-dimensional torus
with a non simply connected group we refer the reader to
\cite{Schweigert:1996tg}.}
\end{example}

Next we can turn to the discussion of the BF-theory on a surface with boundary. Following the logic of
Section \ref{s:group} the boundary conditions of the model are described by admissible branes,
{\it i.e.} submanifolds $\D$ of $\G$ satisfying the properties of Definition
\ref{admissible_branes_definition}. Let us choose a subgroup $\H^\perp \subset \G^*$ with Lie algebra $\h^\perp$:
it is always coisotropic in $\G^*$ since on $\G^*$ Poisson structure is trivial. The complementary dual algebra
$\h$ is defined
as the annihilator of $\h^\perp$.
The corresponding group $\H$ can be identified with $\h$ itself. We can conclude that
$\H$ is a vector subspace of $\G$ which is coisotropic and invariant under the coadjoint action of $\H^\perp$.
  We can construct other admissible branes by considering the fixed points
  of the coadjoint action of $\H^\perp$. Let $g_0$ be a fixed point for $\H^\perp$ then $\D=\H+g_0$ is
still an admissible brane. In more explicit terms let us choose coordinates $X^A=(X^a,X^n)$ on $\G$ such that
$X^a =0$ parametrizes $\H$. Then $\D$ is the hyperplane $X^a=g_0^a$; the coisotropy of $\D$
follows
 \beq
 \alpha^{ab}(X^a=g_0^a, X^\alpha) = \tilde{f}^{ab}_{\,\,\,\,\,\,\,\,\,n} X^n +
 \tilde{f}^{ab}_{\,\,\,\,\,\,\,\,\,c} g_0^c =0 ,
\eeq{coiBFthea}
where $\tilde{f}^{ab}_{\,\,\,\,\,\,\,\,\,n}=0$ since $\h^\perp$ is a subalgebra and
$\tilde{f}^{ab}_{\,\,\,\,\,\,\,\,\,c} g_0^c =0$ since $g_0$ is $\H^\perp$ invariant. It is clear also that these
are the only admissible $(\H,\H^\perp)$ branes.

\begin{example}\label{example_boundary}{\rm
Let us describe these branes more explicitly in the case of $SU(2)$-BF-theory.
We continue the discussion started in Example \ref{example_no_boundary} and we refer now
to the notation of Proposition \ref{p:boundary}. Let us choose $\H^\perp$ as the diagonal
$U(1)$ so that $\H={\bf R}^2=\{(x,y,z=0)\}$. Admissible branes are then $\D(z_0)=\{(x,y,z) \,|\, z=z_0\}$ and
the space $L_{\D(z_0)}$ of $U(1)$-orbits on $\D(z_0)$ is again ${\mathbf R}^+$,
where $\rho\in{\mathbf R}^+$ corresponds to the circle of radius $\rho$ and center $(0,0,z_0)$ in $\D(z_0)$.
If $\rho=0$ then the orbit is made by only a point $x_0=(0,0,z_0)$ and the stability subgroup $\H_{(x_0,\Sy)}$
coincides with $\H^\perp$, {\it i.e.} $p^{-1}(0)$ is just the moduli space of flat $U(1)$ connections over $\Sigma$.
If $\rho>0$ then $\H_{(x_0,\Sy)}\cap\H^\perp=\emptyset$ and we get that $p^{-1}(\rho)$ is the moduli space of
flat $U(1)$ connections over $\Sigma$ such that $J=0$ on the boundary.}
\end{example}

\section{Duality}
\label{s:duality}

In this Section we want to briefly discuss the intriguing duality
properties of the model. It is natural to expect that there may
exist some relation between the Poisson sigma models defined over
the Poisson-Lie group $\G$ and its dual $\G^*$. The nature of this
section is quite speculative, since we cannot offer any definitive
picture.
 Nevertheless we think that it is worthwhile to present the
following observations on the problem.

Let us start by recalling from Section \ref{s:group} the relevant
properties of the model. We have pointed out that the equations of motion of the Poisson sigma model
over the Poisson-Lie group $\G$ can be rewritten as follows
\beq
 d J_A +  \frac{1}{2} \tilde{f}^{CB}_{\,\,\,\,\,\,\,\,\,A}\, J_C \wedge J_B = 0 ,
\eeq{cur1J}
\beq
 dX^\mu - s^{A\mu}(X) J_A =0 .
\eeq{neweqofmot1111}
 The equations (\ref{cur1J})
and (\ref{neweqofmot1111}) are invariant under the following gauge transformations
\beq
 \delta_\beta X^\mu = s^{A\mu} \beta_A,
\eeq{newgautran1aa}
\beq
  \delta_\beta J_A = d \beta_A + \tilde{f}^{CB}_{\,\,\,\,\,\,\,\,\,A} J_C \beta_B .
\eeq{newgautranaa} In Section \ref{s:group} we already drew
attention to the fact that there exists the ``dual'' current
 $j^A$ which satisfies the equation
\beq
 d j^A + \frac{1}{2} f_{BC}^{\,\,\,\,\,\,\,\,\,A} j^B \wedge j^C = 0 ,
\eeq{CartMa123}
which can be interpreted as a Bianchi identity. The equation (\ref{CartMa123}) is invariant
under the gauge transformations
\beq
\delta_\beta j^A =  d\tilde{\beta}^A + f_{CB}^{\,\,\,\,\,\,\,\,\,A} j^C \tilde{\beta}^B .
\eeq{gautrandual1234}

The Poisson sigma model over $\G$ admits a particularly symmetric formulation with the model
defined on $\G^*$. The boundary conditions are formulated in a quite dual way, see all the discussion of Section
\ref{s:group}. Nevertheless we have to stress that the two models
do not have equivalent moduli spaces of solutions.
For example, let us consider the two models defined on the disk. Let us
choose the boundary conditions by fixing the brane to be $\H \subset \G$
and $\H^\perp \subset \G^*$, where $\H$ and $\H^\perp$ are complementary dual coisotropic subgroups.
As we have argued in Example \ref{disk} the corresponding moduli spaces are the space of $\H^\perp$-orbits
on $\H$ and the space of $\H$-orbits on $\H^\perp$.
These two spaces are not related in any obvious way. There is an extreme situation when $\H = \{e\}$
and $\H^\perp = \G^*$  and the moduli spaces  are completely different.
Therefore we conclude that we cannot construct a map between
the solutions of two models taking into account the gauge equivalence. However this observation does not
exclude a more subtle interpretation of the duality for these models.

We should keep in mind that the two currents $J_A$ and $j^A$ (as well as their gauge parameters
$\beta_A$ and $\tilde{\beta}^A$) are not independent in our model.
Let us explore the relations between these two currents and the corresponding gauge transformations.
Using the definitions from Section \ref{s:group} we know
that the currents $j$ and $J$ are related to each other as follows $j^A + \omega^A_\mu s^{B\mu} (X) J_B =0$ and
correspondingly the gauge parameters as $\tilde{\beta}^A - s^{A\mu} \omega^B_\mu (X) \beta_B =0$.
The main goal is to exclude completely $X$ from our consideration and to formulate a new model entirely in terms of
the currents $J$ and $j$ and symmetric in the exchange of $\G$ and $\G^*$.

Using the above relations between the currents we can show that the following properties are trivially satisfied
\beq
 \beta_A \tilde{\beta}^A=0,
\eeq{isotrbeta}
\beq
 J_A \tilde{\beta}^A + j^A \beta_A  =0 ,
\eeq{isotrbeta12}
\beq
 j^A \wedge^* J_A=0 ,
\eeq{isotrbeta1244} where in the last equation we have introduced
a metric on $\Sigma$ in order to have a symmetric pairing of one
forms. We get a perfectly dual system of equations for the two
currents $(j,J)$ defined by the flatness conditions (\ref{cur1J})
and (\ref{CartMa123}) together with (\ref{isotrbeta1244}); the
gauge transformations $(\beta,\tilde{\beta})$ defined by
(\ref{newgautranaa}) and (\ref{gautrandual1234}) must satisfy
(\ref{isotrbeta})-(\ref{isotrbeta12}). In what follows we refer to
this system as the $(j,J)$-system. By construction the Poisson
sigma models over $\G$ and over $\G^*$ can be embedded into this
system of equations. However the $(j,J)$-system admits more
solutions than the original models on $\G$ and on $\G^*$.

Let us analyze the $(j,J)$ system in term of the Drinfeld double ${\mathbf D(g)}=\g\oplus\g^*$.
The two currents $J_A$ and $j^A$ naturally define ${\cal J}=j+J\in\Omega^1\Sigma\otimes{\mathbf D(g)}$
and the gauge transformations  $(\beta_A,\tilde{\beta}^A)$ define ${\cal B}:\Sigma\rightarrow{\mathbf D(g)}$.
Introducing the natural paring $\langle\,\,,\,\,\rangle$ on ${\mathbf D(g)}$ (where for the forms we can use
the Hodge star operation in order to make it symmetric) we can write the conditions
(\ref{isotrbeta})-(\ref{isotrbeta1244}) as follows
\beq
 \langle {\cal B}, {\cal B} \rangle =0,\,\,\,\,\,\,\,\,\,\,
 \langle {\cal J}, {\cal B} \rangle = 0,\,\,\,\,\,\,\,\,\,\,
 \langle {\cal J}, {\cal J} \rangle =  0 .
\eeq{paringdoubls} This conditions are satisfied when ${\cal J}$
and ${\cal B}$ are elements of ${\mathbf l} \subset {\mathbf
D(g)}$, where ${\mathbf l}$ is a maximally isotropic subspace with
respect to the paring on the double.

Let us describe some possible solutions of the $(j,J)$ system. Let us consider
a map $r : {\mathbf g}^* \rightarrow {\mathbf g}$ such that $r^{AB} = - r^{BA}$, or equivalently $r\in\g\wedge\g$.
If we define the currents and the gauge parameters as $j^A= r^{BA}J_{B}$
$\tilde{\beta}^A= r^{BA}\beta_{B}$, then the isotropy conditions (\ref{paringdoubls}) are solved automatically.
In this case the maximally isotropic space ${\mathbf l}$
is defined as follows ${\mathbf l} = \{r(t)+ t\,|\, t\in {\mathbf g}^*\}$. The flatness
condition (\ref{cur1J}) for $J$ implies the flatness condition (\ref{CartMa123}) for $j$
if and only if
$r : {\mathbf g}^* \rightarrow {\mathbf g}$ is a Lie algebra homomorphism. Under the same condition
also the gauge transformations are mapped appropriately and ${\mathbf l}$ is a Lagrangian subalgebra of
the double. For example, one can verify that this situation
can be realized when $r$ solves the classical Yang Baxter equation and the group $\G$ has the corresponding
triangular Poisson-Lie structure.

Another possible solutions of the $(j, J)$-system is given when $j$ is a flat $\H$-connection
and $J$ is a flat $\H^\perp$-connection, where $\H$ and $\H^\perp$ are complementary dual coisotropic subgroups.
Again, if we denote with $\h$ and $\h^\perp$ the corresponding subalgebras, ${\mathbf l}=\h\oplus\h^\perp$ is a
Lagrangian subalgebra of ${\mathbf D(g)}$.

It is tempting to conjecture that a generic solution of the $(j,J)$-system
can be related to Lagrangian subalgebras ${\mathbf l}$ of the double. The solutions
defined by the Poisson sigma model are associated to a fixed symplectic leaf; by the Drinfeld theorem we can
associate to every Poisson homogeneous space (and so to every symplectic leaf) an orbit of Lagrangian subalgebras
of the double \cite{Drinfeld}.

\section{Summary and discussion}
\label{s:end}

In this paper we have presented the analysis of the classical Poisson sigma model
 defined over the Poisson-Lie group $\G$. We have reformulated the on-shell Poisson sigma
 model over $\G$ in terms of the $\G^*$ flat connections $J$ and the parallel sections $X$
 of the associated fibre bundle $\Sigma \times \G$. This reformulation suggests the natural
 description of the boundary conditions which are specified by the coisotropic subgroups.
  Using this description we are able
 to describe the moduli space of the model for the generic compact Riemann surface both
 without and with a boundary. We show that the moduli space is the union of the
 appropriate moduli spaces of the flat connections. At the end of the paper we offer
 our thoughts on the possible relation between the models over $\G$ and $\G^*$ and also
   on the other possible models defined over the whole Drinfeld double.

There are some observations which the presented analysis suggests,
but which we have not
 pursued further in the present paper. One of striking properties is that once we discuss
 the system at the level of the equations of motions we are not confined to two dimensional
 world sheet $\Sigma$. For example, the equations (\ref{eqmotion}) are defined in any dimensions
 and they are invariant under the transformations (\ref{gaugetransf}) provided that $\alpha$ is a Poisson
 structure. In fact the integrability conditions of (\ref{eqmotion}) would require $\alpha$
 to be a Poisson structure. This argument goes in the spirit of the construction used in
 the higher spin theories (e.g., see the discussion in Section 3 of \cite{Vasiliev:sa}). Also we believe
 that the discussion of the moduli spaces in Section \ref{s:moduli} can be generalized
 to arbitrary dimensions. Therefore the Poisson sigma model is defined in any dimensions.
 In more than two dimensions in order to write the action we have to introduce extra
 fields (the Lagrangian multipliers for the equations of motion).
  However if we quantize the system and look at its relevance to the deformation quantization
 then $\dim(\Sigma) =2$.

  Another striking point is the following observation.
 In \cite{GW} it is shown that if  $\G$ is a compact
semisimple Lie group equipped with the standard Poisson-Lie
structure then the the Poisson structure on $\G^*$ is globally
diffeomorphic to the linear structure on $Lie(\G)^*$. It suggests that
 the BF-theory is related to the model over $\G^*$ with the standard Poisson-Lie
 structure. However at present moment it is just a speculation.

 As it was shown in \cite{Cattaneo:1999fm} the Poisson sigma model over the disk
 leads to the Kontsevich star product. When the target is
 a Poisson-Lie group the theory on the disk should be related to quantum
 groups in a certain way. The generic boundary condition on the disk is characterized
 by a coisotropic subgroup as we have described in this paper.
 Infact there is a believe that the coisotropic subgroups survive the quantization \cite{Ciccoli}.
  The coisotropic subgroups define quantum homogeneous spaces for quantum groups and hopefully
 our description of the boundary conditions  in the quantum theory can produce a new insight
 into the subject.

 It is obvious that within this project the next natural step is to consider the
  present model in the context of  the quantum theory.
 We believe that our results should be helpful for the quantization
 of the theory. We are planning to come back to this issue elsewhere.

\bigskip

\bigskip

{\bf Acknowledgements}:
 We are grateful to Nicola Ciccoli, Giorgio
 Ottaviani, Domenico Seminara, Mikhail Vasiliev and Gabriele Vezzosi for discussions and bibliographical
indications. We thank Alexander Popov the e-mail exchange about the flat connections.
 We thank Kevin Graham for the valuable comments on the text.

\appendix

\section{Appendix}
\label{appendix}

 In this Appendix we present the proof that for the (on-shell) general Poisson sigma model
   the image of $X$
 is contained in a single symplectic leaf.
  The similar argument was presented in  \cite{Bojowald:2003pz}.

The first equation in (\ref{eqmotion}) implies
 that  each tangent vector to
$X(\Sigma)$ is tangent to the symplectic leaf $\Sy$. However this fact alone
does not imply that ${\rm Im} X$ is contained in a single leaf. The first equation in (\ref{eqmotion})
 actually
says that the image of each curve $\xi(t)$ on $\Sigma$ is an
integral curve of the (parameter dependent) vector field
$\alpha^{\mu\nu}(X)\eta_{\alpha\nu}(\xi(t))\dot{\xi}^\alpha(t)\partial_\mu$.
Let $x(0)=x_0\in {\cal S}$ then by the splitting theorem (Theorem 2.16 in
\cite{Va}) it exists a neighborhood $U_{x_0}$ which is Poisson
equivalent to ${\cal S}_{x_0}\times N$, where ${\cal
S}_{x_0}={\cal S}\cap U_{x_0}$ and $N$ is a Poisson manifold of
zero rank at $x_0$: the integral curve passing through $x_0$ then is
the direct product of a curve in $\cal S$ and one in $N$. Since
existence and uniqueness of the integral curve is assured also for
parameter dependent vector fields (modulo some assumption, see comment
after Theorem 2.1.2 of \cite{AM}), the integral curve in $N$ passing
by $x_0$ is the constant one.

\end{document}